\documentclass[journal]{IEEEtran}

\ifCLASSINFOpdf
\else
   \usepackage[dvips]{graphicx}
\fi
\usepackage{url}

\usepackage{etoolbox}
\robustify\bfseries
\usepackage{graphicx}
\usepackage{amsmath,amssymb,graphicx}
\usepackage[]{algorithm2e}
\usepackage{tabularx}
\usepackage{wasysym}
\usepackage{color,soul}
\usepackage{cite}
\usepackage{balance}
\usepackage{breqn}
\usepackage{makecell}
\usepackage{booktabs}
\usepackage{multirow}
\usepackage{siunitx}
\usepackage{caption}
\usepackage[table, svgnames]{xcolor}
\usepackage{hyperref}
\def\H{{\mathsf H}}

\def\CC{{\mathbb C}}
\def\RR{{\mathbb R}}
\usepackage{caption}
\usepackage{arydshln}
\usepackage{tabu}

\def\ZQHLrow{}
\newcommand{\ZQHL}[1]{{\color{black}#1}}
\newcommand{\ZQHLSecondRevision}[1]{{\color{black}#1}}
\newcommand{\ZQHLSecondRevisionRedHightligh}[1]{{\color{black}#1}}

\makeatletter
\newcommand*{\bigs}[1]{{\hbox{$\left#1\vbox to9\p@{}\right.\n@space$}}}
\makeatother

\begin{document}

\title{STFT-Domain Neural Speech Enhancement with Very Low Algorithmic Latency}

\author{Zhong-Qiu Wang, Gordon Wichern, Shinji Watanabe, and Jonathan Le Roux
\thanks{Manuscript received on Mar. 14, 2022; revised Aug. 26, 2022; revised Nov. 2, 2022; accepted Nov. 3, 2022.}
\thanks{
Z.-Q. Wang and S. Watanabe are with the Language Technologies Institute, Carnegie Mellon University, Pittsburgh, PA 15213, USA (e-mail: wang.zhongqiu41@gmail.com, shinjiw@cmu.edu).}
\thanks{
G.\ Wichern and J.\ Le Roux are with Mitsubishi Electric Research Laboratories, Cambridge, MA 02139, USA (e-mail: \{wichern,leroux\}@merl.com).}
}

\markboth{}
{Shell \MakeLowercase{\textit{et al.}}: Bare Demo of IEEEtran.cls for IEEE Journals}
\maketitle

\begin{abstract}

Deep learning based speech enhancement in the short-time Fourier transform (STFT) domain typically uses a large window length such as 32 ms. A larger window can lead to higher frequency resolution and potentially better enhancement. This however incurs an algorithmic latency of 32 ms in an online setup, because the overlap-add algorithm used in the inverse STFT (iSTFT) is also performed using the same window size. To reduce this inherent latency, we adapt a conventional dual-window-size approach, where a regular input window size is used for STFT but a shorter output window is used for overlap-add, for STFT-domain deep learning based frame-online speech enhancement. Based on this STFT-iSTFT configuration, we employ complex spectral mapping for frame-online enhancement, where a deep neural network (DNN) is trained to predict the real and imaginary (RI) components of target speech from the mixture RI components. In addition, we use the DNN-predicted RI components to conduct frame-online beamforming, the results of which are used as extra features for a second DNN to perform frame-online post-filtering. The frequency-domain beamformer can be easily integrated with our DNNs and is designed to not incur any algorithmic latency. Additionally, we propose a future-frame prediction technique to further reduce the algorithmic latency. Evaluation on noisy-reverberant speech enhancement shows the effectiveness of the proposed algorithms. Compared with Conv-TasNet, our STFT-domain system can achieve better enhancement performance for a comparable amount of computation, or comparable performance with less computation, maintaining strong performance at an algorithmic latency as low as 2 ms.

\end{abstract}


\IEEEpeerreviewmaketitle

\section{Introduction}

\IEEEPARstart{D}{eep} learning has dramatically advanced speech enhancement in the past decade \cite{WDL2018}.
Early studies estimated target magnitudes via time-frequency (T-F) masking \cite{Y.Wang2013} or directly predicted target magnitude via spectral mapping \cite{Y.Xu2015}, both using the mixture phase for signal re-synthesis.
Subsequent studies strove to improve phase modelling by performing phase estimation via magnitude-driven iterative phase reconstruction \cite{LeRoux2018}.
Recent effort focuses on complex- and time-domain approaches \cite{Williamson2016, Pascual2017, Luo2019, Tan2020, Luo2020,  Subakan2021, Wang2021FCPjournal}, where magnitude and phase are modelled simultaneously through end-to-end optimization.

\begin{figure}
  \centering  
  \includegraphics[width=7cm]{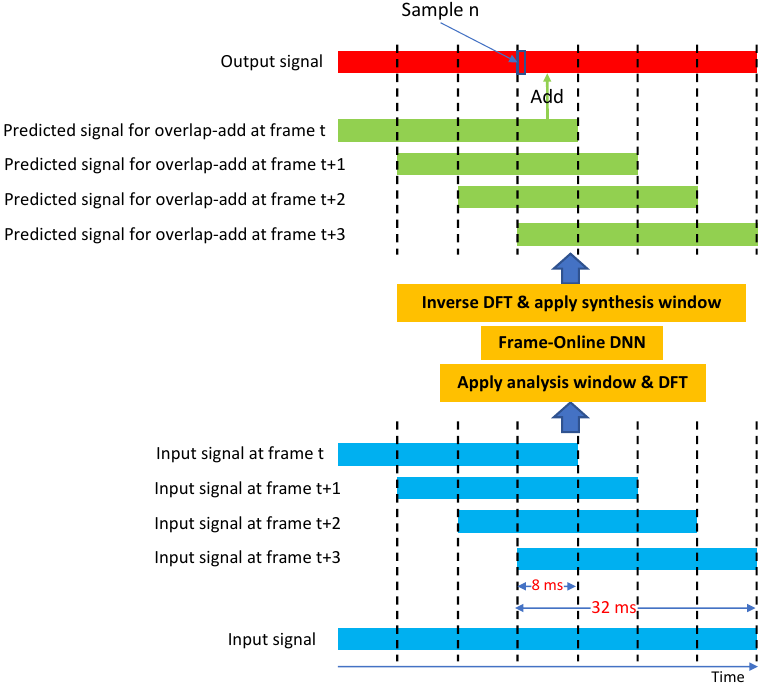}\vspace{-0.1cm} \\  
  \caption{Illustration of processing latency in systems based on regular STFT and iSTFT.
  Each rectangular band denotes a segment of time-domain signals.
  We use 75\% frame overlap as an example.
  Because of the overlap-add in the iSTFT, to get the prediction at sample $n$ (marked in the top of the figure) at frame $t$, one has to first observe all the samples of frame $t+3$ and then wait until the DNN finishes processing frame $t+3$.
  The processing delay is hence the window length plus the running time of processing one frame. %
  }
  \label{overlapregular}\vspace{-0.5cm}
\end{figure}

Many application scenarios such as teleconferencing and hearing aids require low-latency speech enhancement.
Deep learning based approaches \cite{Wilson2018, Tan2020, Pandey2020, Hu2020, Han2020, Tan2021} handle this by using causal DNN blocks, such as uni-directional LSTMs, causal convolutions, causal attention layers, and causal normalization layers.
Although many previous studies along this line advocate that their system with a causal DNN model is causal, one should be aware that most of these systems are, to be more precise, frame-online, and the amount of look-ahead depends on the frame length.
One major approach that can potentially achieve sample-level causal processing is by using WaveNet-like models \cite{Oord2016}.
However, their effectiveness in dealing with noise and reverberation in a sample-causal setup is unclear \cite{Rethage2018}.
In addition, at run time such models need to 
run a forward pass for each sample,
resulting in a humongous and likely unnecessary amount of computation.
Popular STFT- and time-domain approaches typically split signals into overlapped frames with a reasonably large hop length before processing.
One advantage is that the forward pass then only needs to be run
every hop-length samples.
The latency is however equal to the window length due to the use of overlap-add in signal re-synthesis, plus the running time of processing one frame (see Fig.~\ref{overlapregular} and its caption for a detailed explanation of the latency).
In the recent Deep Noise Suppression (DNS) challenge \cite{K.A.Reddy2020}, one key requirement was that the latency of producing an estimate for the sample at index $n$ 
cannot exceed 40 ms on a standard Intel Core i5 processor.
For a typical STFT-based system with a 32 ms window and an 8 ms hop size, a frame-online DNN-based system satisfies the requirement if the processing of each frame can finish within 8 ms on the specified processor. %
We define the latency due to algorithmic reasons (such as overlap-add) as {\bf algorithmic latency}, and the computing time needed to process one frame as {\bf hardware latency}.
The overall latency is the summation of the two and is denoted as {\bf processing latency}.

Although a 40 ms processing latency can meet the demand of many applications, for hearing aids this latency is too large to deliver a good listening experience.
In the recent Clarity challenge, proposed for hearing aid design \cite{ClarityWebpage}, the required algorithmic latency was 5 ms.
Such a low-latency constraint requires new designs and significant modifications to existing enhancement algorithms.
To meet this constraint, our study aims at an enhancement system with a window that looks ahead at most 4 ms of samples, and a 2 ms hop size.
We assume that the hardware latency can be within 2 ms, leading to a maximum processing delay of 6 ms, even though this computational capability might not be available right now for a computationally demanding DNN model on resource-constrained edge devices.
We emphasize that this study focuses on improving enhancement performance and reducing algorithmic latency rather than computational efficiency or feasibility on current hardware.

In the literature, some early STFT-domain beamforming studies \cite{Huang2012, Gode2021} use small window and hop sizes to achieve enhancement with a low algorithmic latency, and a low frequency resolution is used for STFT following the short window length.
However, this low frequency resolution may limit the enhancement performance \cite{WDL2018, Wang2021seq, wangshanshan2021deep} when phase estimation is not performed or the estimated phase is not good enough \cite{Peer2022}.
There are studies \cite{W.Lollmann2008, DItter2020} designing low-delay filterbanks with warped frequencies for speech enhancement and speaker separation, but such manually-designed filterbanks are often complicated and modern deep learning based solutions \cite{WDL2018, Luo2019} tend to learn similar filters through end-to-end training based on the complex T-F representations with uniform frequencies or based on the raw time-domain signals.
In the recent Clarity challenge, almost all the top teams \cite{Tu2021, Zmolikova2021, Gajecki2021} adopt time-domain networks such as Conv-TasNet \cite{Luo2019, ZhangJisi2020}, which can use very short window and hop sizes to potentially realize very low-latency enhancement.
Conv-TasNet leverages DNN-based end-to-end optimization to learn a set of bases for a small window of samples respectively for its encoder and decoder to replace the conventional STFT and iSTFT operations.
The number of bases is set to be much larger than the number of samples in the window.
Enhancement is then performed in the higher-dimensional encoded space and the decoder is used for overlap-add based signal re-synthesis.
While achieving good separation performance in monaural anechoic speaker separation tasks, Conv-TasNet performs less impressively in reverberant conditions and in multi-microphone scenarios than frequency-domain approaches \cite{Heitkaemper2020, Wang2021FCPjournal, Zhang2021}.
In addition, the basis learned by Conv-TasNet is not narrowband \cite{Luo2019}.
It is not straightforward how to combine Conv-TasNet with conventional STFT-domain enhancement algorithms to achieve further gains, without incurring additional algorithmic latency.
Such conventional algorithms include beamforming and weighted prediction error (WPE), which rely on the narrow-band assumption and can produce reliable separation through their per-frequency processing \cite{Yoshioka2012, Haeb-Umbach2020, Wang2021FCPjournal, Gannot2017}.
One way of combining them \cite{Ochiai2020, Wang2021seq, Chen2021} is by iterating Conv-TasNet, which uses a very short window, with STFT-domain beamforming, which uses a regular, longer window.
To use Conv-TasNet outputs to compute signal statistics for STFT-domain beamforming, one has to first re-synthesize time-domain signals before extracting STFT spectra for beamforming.
Similarly, to apply Conv-TasNet on beamforming results for post-filtering, one has to apply iSTFT to get time-domain signals before feeding them to Conv-TasNet.
Such an iterative procedure would however gradually build up the algorithmic latency, because the overlap-add algorithms are used multiple times in Conv-TasNet and iSTFT.

\begin{figure}
  \centering  
  \includegraphics[width=9cm]{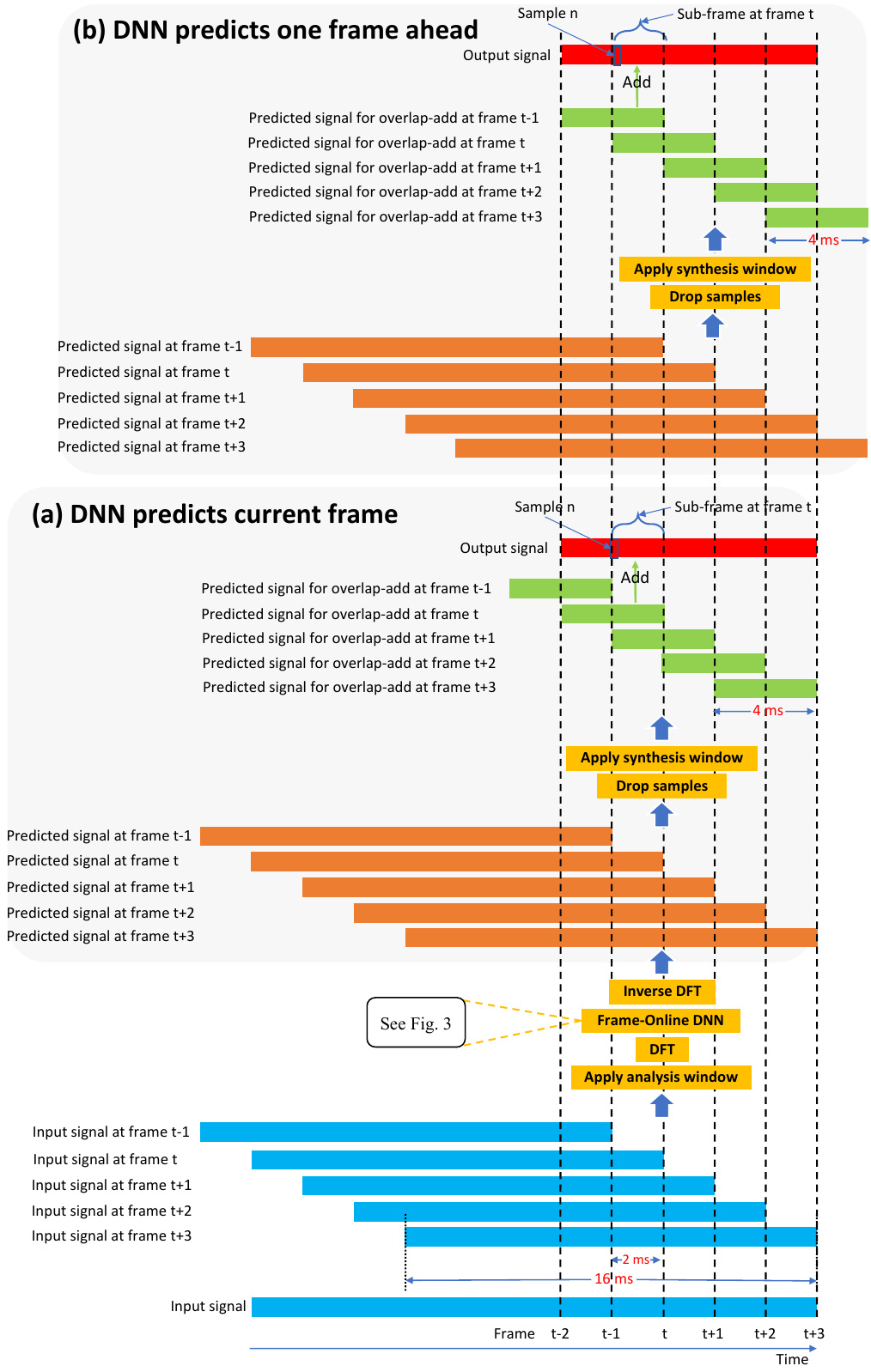} \vspace{-0.4cm} \\  
  \caption{
  Illustration of overlap-add with dual window sizes. This example uses a 16 ms input window size for STFT, a 4 ms output window size for overlap-add, and a 2 ms hop size. \ZQHL{At each frame, DNN is trained to predict (a) current frame; (b) one frame ahead.}
  }
  \label{overlapdual}\vspace{-0.5cm}
\end{figure}

It is commonly perceived that regular STFT-based systems, which suffer from a large algorithmic latency equal to the STFT's typically long window length, are not ideal for very low-latency speech enhancement, and time-domain models such as Conv-TasNet, which can achieve strong performance using very short windows, appear a more appropriate choice \cite{Luo2019}.
In this study, we show that our STFT-based system can also produce a comparable or better enhancement performance at an algorithmic latency as low as 4 or 2 ms.
This is partially achieved by combining STFT-domain, deep learning based speech enhancement with a conventional dual window approach \cite{Mauler2007, Wood2019}, which uses a regularly long window length for STFT and a shorter window length for overlap-add.
This approach is illustrated in Fig.~\ref{overlapdual}(a).
More specifically, assuming a hop size (HS) of 2 ms, we use a 16 ms input window size (iWS) for STFT, and an output window size (oWS) of 4 ms for the overlap-add in iSTFT.
The 16 ms input window looks 4 ms ahead and 12 ms in the past in this case.
After obtaining the predicted signal at each frame, i.e., after performing inverse discrete Fourier transform (iDFT), we throw away the first 12 ms of waveforms, apply a synthesis window, and perform overlap-add based on the last 4 ms of signals at each frame.
The DNN module is designed to be frame-online. Therefore the entire system has an algorithmic latency of 4 ms.
Later in Section~\ref{futureprediction}, we will introduce a future-frame prediction technique to further reduce the algorithmic latency to 2 ms.

When used with DNNs, this dual window size approach has several advantages.
First, using a
long window for STFT leads to higher frequency resolution, meaning that we could have more estimated filters (or mask values) per frame to obtain more fine-grained enhancement.
In addition, higher frequency resolution could better leverage the speech sparsity property in the T-F domain for enhancement \cite{WDL2018, wangshanshan2021deep}.
Second, using a longer input window can capture more reverberation at each frame, potentially leading to better dereverberation.
In addition, it could lead to better spatial processing, as the inter-channel phase patterns could be more stable and salient for longer signals.
Third, STFT bases are narrowband in nature, meaning that we can readily use our DNN outputs (in this study, the estimated target real and imaginary components) to compute a conventional frequency-domain beamformer, whose results can be used as extra features for another DNN to better predict the target speech.

\ZQHL{
The major contributions of this paper are provided below, together with our justifications:

First, we adapt a conventional dual window size approach \cite{Mauler2007, Wood2019} to reduce the algorithmic latency of STFT-domain deep learning based speech enhancement.
Although using a synthesis window shorter than the analysis window has been proposed in conventional non-deep-learning speech enhancement studies \cite{Mauler2007, Wood2019}, it is seldomly employed (or investigated) in modern deep learning based speech enhancement.
TasNet-style time-domain DNN models \cite{Luo2019, Luo2020}, which learn encoders (i.e., filterbanks) and decoders based on very short windows of signals through end-to-end training, have become the dominant approach to achieve very low-latency enhancement.
This can be observed from the top solutions \cite{Tu2021, Zmolikova2021, Gajecki2021} in the recent Clarity challenge \cite{ClarityWebpage}.
Such success casts doubt on whether STFT-domain approaches are inherently sub-optimal compared with time-domain approaches, and whether the go-to approach for very low-latency speech enhancement should be time-domain approaches.
In our experiments, we find that the proposed STFT-domain system can achieve comparably good or better performance than popular Conv-TasNet systems \cite{Luo2019, ZhangJisi2020, Tu2021}, using a similar amount of computation and at an algorithmic latency as low as 4 or 2 ms.
To the best of our knowledge, to date there is only one deep learning based study exploring the dual window size idea \cite{wangshanshan2021deep}.
However, that work focuses on a speaker separation task rather than speech enhancement, only tackles the monaural condition, performs real-valued masking based on a weak LSTM model, only reduces the algorithmic latency to 8 ms, and does not perform a comparison with time-domain models.
Therefore, whether one should embrace STFT-domain approaches for very low-latency speech enhancement remains unclear.
In contrast, our study includes both single- and multi-channel conditions, considers both magnitude and phase estimation through complex spectral mapping based on modern DNN architectures, reduces the algorithmic latency to as low as 2 ms (where our models still show competitive performance), and, most importantly, we perform a thorough comparison with the representative time-domain model, Conv-TasNet.

Second, we utilize the outputs from the first DNN for frequency-domain frame-online beamforming, and the beamforming result is fed to a second DNN for better enhancement (i.e., post-filtering).
This beamforming followed by post-filtering approach produces clear improvements in our experiments over just using one DNN (i.e., not using any beamforming and post-filtering).
Although this approach has been studied in non-causal neural speech enhancement~\cite{Wang2020d, Wang2020chime, Wang2020css, Wang2021FCPjournal}, one important advantage we will demonstrate in this paper is that, since the two DNNs and the beamformer all operate in the complex T-F domain, this approach does not incur additional algorithmic latency.
In contrast, time-domain models cannot be straightforwardly combined with frequency-domain beamforming without incurring extra algorithmic latency.
This comparison demonstrates that one advantage of performing very low-latency enhancement in the STFT domain is the integration with frequency-domain beamforming.

Third, we propose a future-frame prediction technique that can further reduce the algorithmic latency caused by the output window size.
We show that predicting one frame ahead only slightly degrades the performance.
This is a significant contribution, because it suggests a good way to reduce the algorithmic latency using the same amount of computation, or maintain the same algorithmic latency but use less computation.
In addition, we analyze the effects of the shape of analysis windows on future-frame prediction, and present preliminary results of predicting multiple frames ahead, which can potentially reduce the algorithmic as well as processing latency to zero.
Furthermore, this future-frame prediction technique is particularly helpful to the two-DNN system with an intermediate beamformer.
We point out that one inherent weakness of the two-DNN system is that, since it stacks two DNNs, the amount of computation would at least be doubled if the same DNN architecture is used in both networks.
Supposing the overlap between output windows is 50\% (for example by using 2 ms oWS and 1 ms hop), we propose to cut the doubled amount of computation by approximately half via doubling the hop size and predicting one frame ahead, at the same time maintaining the same algorithmic latency.
The resulting system still shows competitive performance to Conv-TasNet.
}

\ZQHLSecondRevision{
It should be noted that conventional signal-processing-based acoustic echo control and active noise control studies have shown that low-delay time-domain filtering can be achieved by T-F domain processing~\cite{W.Lollmann2008}, and the idea of predicting future samples (via for example multi-step linear prediction) to reduce processing latency has been studied in active noise control~\cite{Iotov2022}. They are however seldom studied in the context of machine-learning-based speech processing.
}

The following section gives an overview of our system.

\section{System Overview}\label{systemoverview}

Given an utterance of a speaker recorded in noisy-reverberant conditions by a $P$-microphone array, the physical model in the STFT domain can be formulated as
\begin{align} 
	\mathbf{Y}(t,f) &= \mathbf{X}(t,f)+\mathbf{V}(t,f) \nonumber \\
	&= \mathbf{S}(t,f)+\mathbf{H}(t,f)+\mathbf{V}(t,f), \label{eq:phymodel_freq}
\end{align}
where $\mathbf{Y}(t,f)$, $\mathbf{V}(t,f)$, $\mathbf{X}(t,f)$, $\mathbf{S}(t,f)$ and $\mathbf{H}(t,f)\in \CC^{P}$ respectively denote the STFT vectors of the mixure, reverberant noise, reverberant target speech, direct-path and non-direct signals of the target speaker at time $t$ and frequency $f$.
In the rest of this paper, when dropping $t$ and $f$ from the notation, we refer to the corresponding spectrogram.
In our experiments, the default iWS, oWS, and HS for STFT are respectively set to 16, 4, and 2 ms, and the sampling rate is 16 kHz. 
A 256-point DFT is applied to extract 129-dimensional STFT coefficients at each frame.
The analysis window and synthesis window will be described in Section~\ref{window}.

Based on the input $\mathbf{Y}$, we aim at recovering the target speaker's direct-path signal captured at a reference microphone $q$, i.e., $S_q$.
We use the corresponding time-domain signal of $S_q$, denoted as $s_q$, as the reference signal for metric computation.
Note that early reflections are not considered as part of target speech.
In multi-microphone cases, we assume that the same array geometry is used for training and testing, following \cite{Wang2020d, Wang2020css}.
This is a valid assumption as real-world products such as 
smart speakers have a fixed array configuration.

\begin{figure}
  \centering  
  \includegraphics[width=8.5cm]{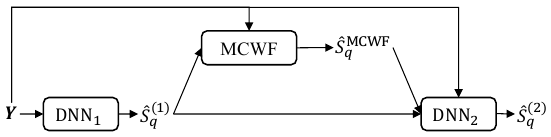}\vspace{-0.1cm}
  \caption{DNN overview. Our system contains two frame-online DNNs with a frame-online multi-channel Wiener filter (MCWF) in between.}
  \label{system_overview_figure} \vspace{-0.5cm}
\end{figure}

Our best performing system, illustrated in Fig.~\ref{system_overview_figure}, has two DNNs.
Using the real and imaginary (RI) components of multiple input signals as input features, the DNNs are trained sequentially based on single- or multi-microphone complex spectral mapping \cite{Williamson2016, Wang2020d, Wang2020css} to predict the RI components of $S_q$.
The estimated speech by DNN$_1$ is used to compute, at each frequency, a multi-channel Wiener filter (MCWF) \cite{Wang2021seq} for the target speaker.
$\text{DNN}_2$ concatenates the RI components of the beamforming results, the outputs of $\text{DNN}_1$, and $\mathbf{Y}$ as features to further estimate the RI components of $S_q$.
The $\text{DNN}_1$, $\text{DNN}_2$, and MCWF modules are all designed to be frame-online, so that we can readily plug our two-DNN system into Fig.~\ref{overlapdual} to achieve enhancement with very low algorithmic latency.
Note that such two-DNN systems with a beamformer in between have been explored in our previous studies \cite{Wang2020d, Wang2020chime, Wang2020css, Wang2021FCPjournal}, but their target was offline processing. This paper extends them for frame-online processing with a very low algorithmic latency.

The rest of this paper is organized as follows. Section~\ref{dnndescription} details the DNN configurations, Section~\ref{fcpandmcwf} describes the DNN-supported beamforming, and Section~\ref{lowlatency} presents the proposed enhancement system with low algorithmic latency.
Experimental setup and evaluation results are presented in Sections~\ref{experiments} and \ref{results}.
Section \ref{conclusion} concludes this paper.

\section{DNN Configurations}\label{dnndescription}

Our DNNs are trained to do complex spectral mapping \cite{Williamson2016}, where the real and imaginary (RI) components of multiple signals are concatenated as input for DNNs to predict the target RI components at a reference microphone.
This section describes the loss functions and the DNN architectures.
The key differences from our earlier studies \cite{Wang2020d, Wang2020chime, Wang2020css, Wang2021FCPjournal} include the facts that (1) we train through the dual window size approach and define the loss function on the re-synthesized signals; and (2) we dramatically reduce the per-frame amount of computation of the DNN models used in our earlier studies.

\subsection{Loss Functions}

The two DNNs in Fig.~\ref{system_overview_figure} are trained using different loss functions.
For DNN$_1$, following \cite{Wang2021compensation, Wang2020d, Wang2020css} the loss is defined on the predicted RI components and their magnitude:
\begin{align}\label{ri+mag}
\mathcal{L}_{\text{RI+Mag}} =
\| \hat{R}_q^{(1)} - \text{Real}&(S_q)\|_1 + \| \hat{I}_q^{(1)} - \text{Imag}(S_q)\|_1 \nonumber \\
&+ \Big\| \sqrt{\hat{R}_q^{(1)^2}+\hat{I}_q^{(1)^2}} - |S_q|\Big\|_1,
\end{align}
where $\hat{R}_q^{(1)}$ and $\hat{I}_q^{(1)}$ are the predicted RI components by DNN$_1$, $\text{Real}(\cdot)$ and $\text{Imag}(\cdot)$ extract RI components, and $\| \cdot\|_1$ computes the $L_1$ norm.
The estimated target spectrogram at the reference microphone $q$ is $\hat{S}_q^{(1)}=\hat{R}_q^{(1)}+j\hat{I}_q^{(1)}$, where $j$ denotes the imaginary unit.

Given the predicted RI components $\hat{R}_q^{(2)}$ and $\hat{I}_q^{(2)}$ by DNN$_2$, we denote $\hat{S}_q^{(2)}=\hat{R}_q^{(2)}+j\hat{I}_q^{(2)}$ and compute the re-synthesized signal $\hat{s}_q^{(2)}=\text{iSTFT}(\hat{S}_q^{(2)})$, where $\text{iSTFT}(\cdot)$ uses a shorter output window for overlap-add to reduce the algorithmic latency (see Fig.~\ref{overlapdual})(a).
The loss function is then defined on the re-synthesized time-domain signal and its STFT magnitude:
\begin{align}\label{wav+mag}
\mathcal{L}_{\text{Wav+Mag}} =
\| \hat{s}_q^{(2)} - &s_q \|_1 \nonumber \\
&+ \Big\| |\text{STFT}_{\mathcal{L}}(\hat{s}_q^{(2)})| - |\text{STFT}_{\mathcal{L}}(s_q)| \Big\|_1,
\end{align}
where $\text{STFT}_{\mathcal{L}}(\cdot)$ extracts a complex spectrogram.
The loss on magnitude is found to consistently improve objective metrics such as PESQ and STOI \cite{Wang2021compensation}.
Note that $\text{STFT}_{\mathcal{L}}(\cdot)$ here can use any window types and window and hop sizes, and can be different from the ones we use to extract $\mathbf{Y}$ and $S_q$, since it is only used for loss computation.
In our experiments, we use the square-root Hann window, a 32 ms window size and an 8 ms hop size to compute this magnitude loss.
Please do not confuse these STFT parameters with the STFT parameters we used to extract $\mathbf{Y}$ and $S_q$.

In our experiments, we will compare the two-DNN system with a %
single-DNN system (i.e., without the beamforming module and the second DNN). In the single-DNN case, DNN$_1$ can be trained using either Eq.~(\ref{ri+mag}) or (\ref{wav+mag}).
Differently, in the two-DNN case, DNN$_1$ is trained only using Eq.~(\ref{ri+mag}).
This is because the beamformer we will derive later in Eq.~(\ref{mcwf}) is based on a DNN-estimated target in the complex domain.

\begin{figure}
  \centering  
  \includegraphics[width=8.5cm]{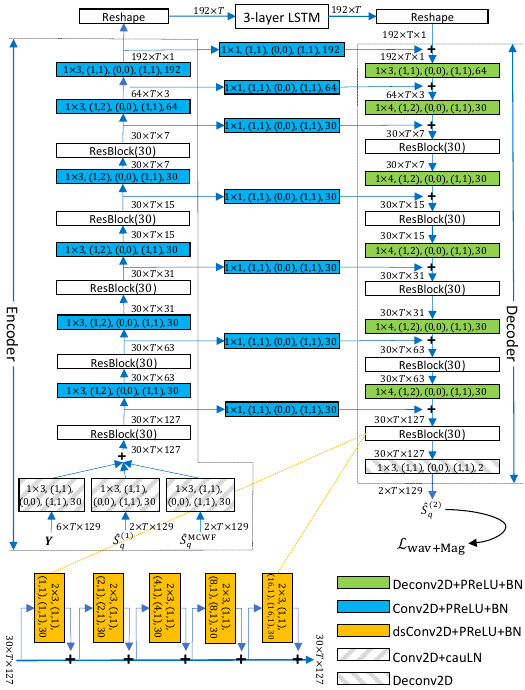} 
  \caption{
  Network architecture of $\text{DNN}_2$.
  Each one of Conv2D, Deconv2D, Conv2D+PReLU+BN, dsConv2D+PReLU+BN, and Deconv2D+PReLU+BN blocks is specified in the format: \textit{kernelTime$\times$kernelFreq, (strideTime, strideFreq), (padTime, padFrequency), (dilationTime, dilationFreq), featureMaps}.
  During training, the tensor shape after each block in the encoder and decoder is denoted in the format:
  \textit{featureMaps$\times$timeSteps$\times$freqChannels}.
  Best viewed in color.
  }
  \label{dnnfigure}\vspace{-0.5cm}
\end{figure}

\subsection{Network Architecture}\label{dnn}

Our DNN architecture, denoted as LSTM-ResUNet, is illustrated in Fig.~\ref{dnnfigure}.
It is a long short-term memory (LSTM) network clamped by a U-Net \cite{Ronneberger2015}.
Residual blocks are inserted at multiple frequency scales in the encoder and decoder of the U-Net.
The motivation of this network design is that U-Net can maintain fine-grained local structure via its skip connections and model contextual information along frequency through down- and up-sampling, LSTM can leverage long-range information, and residual blocks can improve discriminability.
We stack the RI components of different input and output signals as features maps in the network input and output.
DNN$_1$ and DNN$_2$ differ only in their network input.
DNN$_1$ uses the RI components of $\mathbf{Y}$ to predict the RI components of $S_q$, and DNN$_2$ additionally uses as input the RI components of $\hat{S}_q^{(1)}$ and a beamforming result $\hat{S}_q^{\text{MCWF}}$, which will be described later in Section~\ref{fcpandmcwf}.
The encoder contains one two-dimensional (2D) convolution followed by causal layer normalization (cauLN) for each input signal, and six convolutional blocks, each with 2D convolution, parametric ReLU (PReLU) non-linearity, and batch normalization (BN), for down-sampling.
The LSTM contains three layers, each with 300 units.
The decoder includes six blocks of 2D deconvolution, PReLU, and BN, and one 2D deconvolution, for up-sampling.
Each residual block in the encoder and decoder contains five depth-wise separable 2D convolution (denoted as dsConv2D) blocks, where the dilation rate along time are respectively 1, 2, 4, 8 and 16.
Linear activation is used in the output layer to obtain the predicted RI components.

All the convolution and normalization layers are causal (i.e., frame-online) at run time.
We use $1\times3$ or $1\times4$ kernels along time and frequency for the down- and up-sampling convolutions, following \cite{Tan2021}.
Causal $2\times3$ convolutions are used in the residual blocks, following \cite{Tan2021}.

Note that this architecture is similar to the earlier TCN-DenseUNet architecture \cite{Wang2020d, Wang2020chime, Wang2020css, Wang2021FCPjournal}.
Major changes include replacing the DenseNet blocks with residual blocks and replacing regular 2D convolutions with depthwise separable 2D convolutions.
These changes dramatically reduce the amount of computation and the number of trainable parameters.
The network contains around 2.3 million parameters, \ZQHLSecondRevision{compared with the 6.9 million parameters in TCN-DenseUNet.}
It uses similar amount of computation compared with the Conv-TasNet used in our experiments.

\section{Frequency-Domain Beamforming}\label{fcpandmcwf}

Based on the DNN-estimated target RI components, we compute an online multi-channel Wiener filter (MCWF) \cite{Gannot2017} to enhance target speech (see Fig.~\ref{system_overview_figure}).
The MCWF is computed per frequency, leveraging the narrow-band property of STFT.
Although the beamforming result of such a filter usually does not show better scores in terms of enhancement metrics than the immediate DNN outputs, it can provide complementary information to help DNN$_2$ obtain better enhancement results \cite{Wang2020css, Wang2021FCPjournal, Wang2021LowDistortion}.
\ZQHLSecondRevisionRedHightligh{Our main contribution here is to show that frame-online frequency-domain beamforming can be easily integrated with our STFT-domain DNNs to improve enhancement, while not incurring any algorithmic latency.
We can use more advanced beamformers, or dereverberation algorithms such as WPE \cite{Yoshioka2012, Haeb-Umbach2020, Wang2021FCPjournal}, to achieve even better enhancement than MCWF.
They are however out of the scope of this study.}
This section will first describe the offline time-invariant implementation of the MCWF beamformer and then extend it to frame-online.

The MCWF \cite{Gannot2017} computes a linear filter per T-F unit or per frequency to project the mixture to target speech.
Assuming the target speaker does not move within each utterance and based on the DNN-estimated target speech $\hat{S}_q^{(1)}$, we compute a time-invariant MCWF per frequency through the following minimization problem:
\begin{align}\label{mcwf}
\underset{\mathbf{w}(f;q)}{{\text{min}}} 
\sum\nolimits_t \big|
\hat{S}_q^{(1)}(t,f) - \mathbf{w}(f;q)^{\H} \mathbf{Y}(t,f)
\big|^2,
\end{align}
where $q$ denotes the reference microphone and $\mathbf{w}(f;q)\in \CC^{P}$ is a $P$-dimensional filter.
Since the objective is quadratic, a closed-form solution is available:
\begin{align}\label{mcwfcov}
\hat{\mathbf{w}}(f;q) &= \Big(\hat{\mathbf{\Phi}}^{(yy)}(f)\Big)^{-1} \hat{\mathbf{\Phi}}^{(ys)}(f)\mathbf{u}_q, \\
\hat{\mathbf{\Phi}}^{(yy)}(f) &= \sum\nolimits_t \mathbf{Y}(t,f) \mathbf{Y}(t,f)^{\H}, \\
\hat{\mathbf{\Phi}}^{(ys)}(f) &= \sum\nolimits_t \mathbf{Y}(t,f) \hat{\mathbf{S}}^{(1)}(t,f)^{\H}, \label{mcwfcovend}
\end{align}
where $\hat{\mathbf{\Phi}}^{(yy)}(f)$ denotes the observed mixture spatial covariance matrix, $\hat{\mathbf{\Phi}}^{(ys)}(f)$ the estimated covariance matrix between the mixture and the target speaker, and $\mathbf{u}_q$ is a one-hot vector with element $q$ equal to one.
Notice that we do not need to first fully compute the matrix $\hat{\mathbf{\Phi}}^{(ys)}(f)$ and then take its $q^{\text{th}}$ column by multiplying it with $\mathbf{u}_q$, because 
\begin{align}\label{matcol}
\hat{\mathbf{\Phi}}^{(ys)}(f)\mathbf{u}_q = \sum\nolimits_t \mathbf{Y}(t,f) \Big(\hat{S}_q^{(1)}(t,f)\Big)^{*},
\end{align}
where $(\cdot)^{*}$ computes complex conjugate.
The beamforming result is computed as 
\begin{align}\label{mcwfresult}
\hat{S}_q^{\text{MCWF}}(t,f) = \hat{\mathbf{w}}(f;q)^{\H}\mathbf{Y}(t,f).
\end{align}

We point out that, in Eq.~(\ref{mcwf}), the DNN-estimated magnitude and phase can both be used for computing the MCWF beamformer, \ZQHL{while previous studies use DNN-estimated real-valued magnitude masks to compute spatial covariance matrices and then derive MCWF beamformers~\cite{Wang2021seq}.
In addition, we only need to have DNN$_1$ to estimate the target speech at the reference microphone $q$ to compute the MCWF.
Differently, earlier studies perform minimum variance distortionless response (MVDR) beamforming in this two-DNN approach~\cite{Wang2020, Wang2020chime, Wang2020css}, and, to leverage both DNN-estimated magnitude and phase to compute the MVDR beamformer}, they estimate the target speech at all the microphones by training a multi-channel input and multi-channel output network that can predict the target speech at all the microphones at once \cite{Wang2020d, Han2020}, or by running a well-trained multi-channel input and single-channel output network $P$ times at inference time \cite{Wang2020css, Zhang2021}, where each microphone is considered as the reference microphone in turn.
However, the former approach produces worse separation at each microphone than the latter, probably because there are many more signals to predict \cite{Wang2020d}, and the latter dramatically increases the amount of computation \cite{Wang2020css}.
\ZQHL{By using the MCWF in Eq.~(\ref{mcwf}) instead of an MVDR beamformer, we can simplify the beamforming module, as (\ref{mcwf}) just performs a linear projection and does not require an estimated steering vector, and, in addition, we only need DNN$_1$ to estimate the target speech at a reference microphone in order to use both DNN-estimated magnitude and phase for beamformer computation.}

Differently from Eqs.~(\ref{mcwfcov})-(\ref{mcwfcovend}), in a frame-online setup the statistics are accumulated online, similarly to \cite{Higuchi2018}, and the beamformer at each time step is computed as
\begin{align}\label{onlinemcwf}
\hat{\mathbf{w}}(t,f;q) &= \Big(\hat{\mathbf{\Phi}}^{(yy)}(t,f)\Big)^{-1} \hat{\mathbf{\Phi}}^{(ys)}(t,f)\mathbf{u}_q, \\
\hat{\mathbf{\Phi}}^{(yy)}(t,f) &= \hat{\mathbf{\Phi}}^{(yy)}(t-1,f) + \mathbf{Y}(t,f) \mathbf{Y}(t,f)^{\H}, \\
\hat{\mathbf{\Phi}}^{(ys)}(t,f) &= \hat{\mathbf{\Phi}}^{(ys)}(t-1,f) + \mathbf{Y}(t,f) \hat{\mathbf{S}}^{(1)}(t,f)^{\H},
\end{align}
with $\hat{\mathbf{\Phi}}^{(yy)}(0,f)$ and $\hat{\mathbf{\Phi}}^{(ys)}(0,f)$ initialized to be all-zero.
\ZQHL{Note that here we do not use the typical recursive averaging technique~\cite{Gannot2017}, as the target speaker and non-target sources are assumed not to be moving within each utterance.
We will explore the idea of recursive averaging in future work.}

Based on the online time-varying filter $\hat{\mathbf{w}}(t,f;q)$, the beamforming result is obtained as 
\begin{align}\label{onlinemcwfresult}
\hat{S}_q^{\text{MCWF}}(t,f) = \hat{\mathbf{w}}(t,f;q)^{\H}\mathbf{Y}(t,f).
\end{align}

Similarly to \cite{Higuchi2018}, in a frame-online setup $\hat{\mathbf{\Phi}}^{(yy)}(t)^{-1}$ in Eq.~(\ref{onlinemcwf}) can be computed iteratively according to the Woodbury formula, i.e.,
\begin{align}\label{woodbury}
\hat{\mathbf{\Phi}}^{(yy)}(t)^{-1} &= \Big(\hat{\mathbf{\Phi}}^{(yy)}(t-1) + \mathbf{Y}(t) \mathbf{Y}(t)^{\H}\Big)^{-1} \nonumber \\
&\hspace{-1.5cm}= \hat{\mathbf{\Phi}}^{(yy)}(t-1)^{-1} \nonumber \\
&\hspace{-.8cm}- \frac{\hat{\mathbf{\Phi}}^{(yy)}(t-1)^{-1}
\mathbf{Y}(t)
\mathbf{Y}(t)^{\H}
\hat{\mathbf{\Phi}}^{(yy)}(t-1)^{-1}}
{1+\mathbf{Y}(t)^{\H} \hat{\mathbf{\Phi}}^{(yy)}(t-1)^{-1} \mathbf{Y}(t)},
\end{align}
where the frequency index $f$ is dropped to make the equation less cluttered.
This way, expensive matrix inversion at each T-F unit is avoided in the frame-online case.

\section{Enhancement with Low Algorithmic Latency}\label{lowlatency}

In Fig.~\ref{system_overview_figure}, there are two DNNs and an MCWF in between.
Since the DNNs and beamformer all operate in the complex T-F domain, without going back and forth to the time domain, we can use the same STFT resolution for all of them to obtain a two-DNN system with a low algorithmic latency.
This is different from earlier studies \cite{Ochiai2020, Wang2021seq, Chen2021} that combine time-domain models with beamforming and have to switch back and forth to the time domain.
Given a small hop size (say 2 ms), we can use a regular, large iWS (for example 16 ms) for STFT to have a reasonably high frequency resolution for frequency-domain beamforming.
To re-synthesize $\hat{S}_q^{(2)}$ to a time-domain signal, we use the last 4 ms of the 16 ms signals produced by iDFT at each frame for overlap-add, following the procedure illustrated in Fig.~\ref{overlapdual}(a).
The resulting system has an algorithmic latency of 4 ms, even though the STFT spectrograms are extracted using a window size of 16 ms.
The rest of this section describes the analysis and synthesis windows, and a future-frame prediction technique that can further reduce the 4 ms algorithmic latency to 2 ms.

\subsection{Analysis and Synthesis Window Design}\label{window}

\begin{figure}
  \centering  
  \includegraphics[width=6cm]{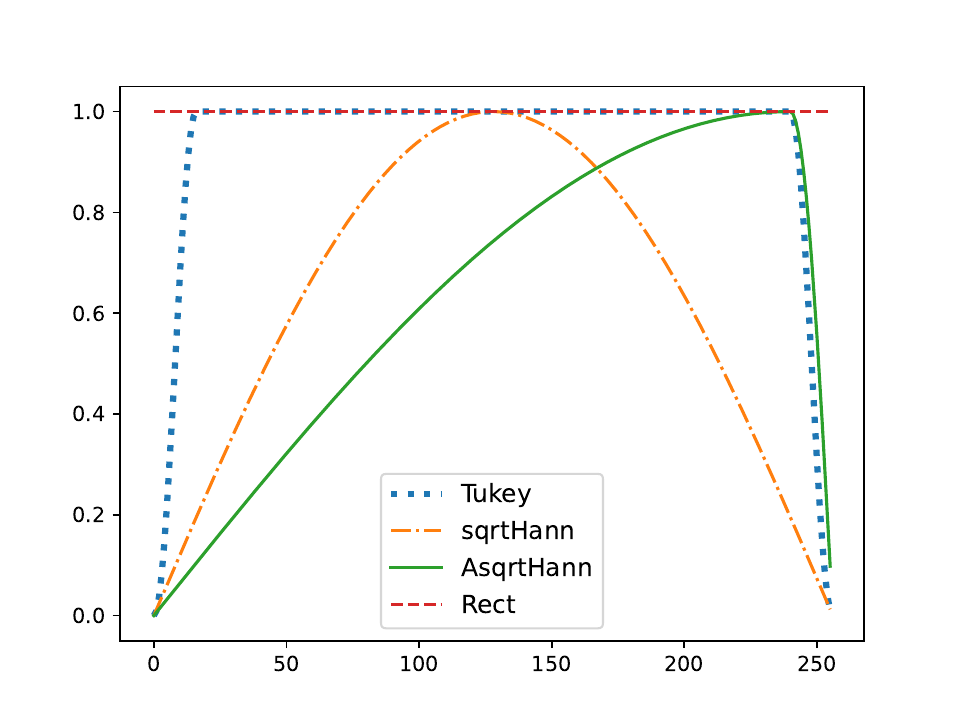} 
  \caption{
  Illustration of analysis windows (assuming a window size of 16 ms and a sampling rate of 16 kHz).
  }
  \label{awinfigure}\vspace{-0.5cm}
\end{figure}

In our experiments, we will investigate various analysis windows such as the square-root Hann (sqrtHann) window, rectangular (Rect) window, asymmetric sqrtHann (AsqrtHann) window \cite{Mauler2007, Wood2019, wangshanshan2021deep}, and Tukey window \cite{Harris1978}.
See Fig.~\ref{awinfigure} for an illustration of the windows.
Our consideration for this investigation is that for windows such as the sqrtHann window, where all the last 4 ms are in the tapering range, the tapering could make the extracted frequency components in the STFT spectrograms less representative of the last 4 ms of signals, where we aim to make predictions.
One solution is to use a rectangular analysis window, which does not taper samples.
However, it is well-known that rectangular windows lead to more spectral leakage due to their higher sidelobes than windows with a tapering shape \cite{Oppenheim, Harris1978}.
Such leakage could degrade per-frequency beamforming as well as the performance of DNNs.
On the other hand, the rectangular window does not taper any samples in the right end and hence the signal in that region is not modified by the window. This could be helpful for very low-latency processing, as we need to make predictions for these samples.
Our study also considers the Tukey window \cite{Harris1978}, defined as
\begin{align}\label{tukeywindow}
g[n] &= 0.5 - 0.5\,\text{cos}\Big(\frac{\pi n}{\alpha N}\Big), &&\text{if\,\,\,} 0 \leq n \leq \alpha N; \nonumber \\
&= g[N-n], &&\text{if\,\,\,} N - \alpha N \leq n < N; \nonumber \\
&= 1, &&\text{otherwise};
\end{align}
where $0 \leq n < N$ and tapering only happens at the first and the last $\alpha N$ samples.
Given a 16 ms analysis window, we set $\alpha$ to $\frac{1}{16}$, meaning 1 ms of tapering on both ends.
We also consider the AsqrtHann window proposed in \cite{Mauler2007, Wood2019, wangshanshan2021deep}, and construct a 16 ms long AsqrtHann window by combining the first half of a 30 ms sqrtHann window and the second half of a 2 ms sqrtHann window.

We compute a synthesis window that can achieve perfect reconstruction when used with an analysis window $g$, following \cite{Griffin1984a}.
Suppose that the run-time oWS in samples is $A$ and the hop size in samples is $B$, and that $A$ is a multiple of $B$, we obtain the synthesis window $l \in \RR^A$ based on the last $A$ samples of the analysis window:
\begin{align}\label{tukeywindowsynthesis}
l[n] &= \frac{g[N-A+n]}{\sum_{k=0}^{A/B-1} {g[N-A+(n \text{\,\,mod\,\,} B) + kB]}^2},
\end{align}
where $0 \leq n < A$.

\subsection{Future-Frame Prediction}\label{futureprediction}

We can further reduce the algorithmic latency by training the DNN to predict, say, one frame ahead.
That is, at time $t$, the DNN predicts the target RI components at frame $t+1$.
This can reduce the algorithmic latency from 4 to 2 ms.
We use Fig.~\ref{overlapdual}(b) to explain the idea.
To get the prediction at sample $n$ (see the top of Fig.~\ref{overlapdual}(b)), we need to overlap-add the last 4 ms of frame $t$ and $t+1$.
If the DNN only predicts the current frame, we can only do the overlap-add after we fully observe frame $t+1$ and finish feed-forwarding frame $t+1$.
In contrast, if the DNN predicts frame $t+1$ at frame $t$, we can do the overlap-add after we observe and finish feed-forwarding frame $t$.
The algorithmic latency is hence reduced by 2 ms.
We can predict more frames ahead to reduce the algorithmic latency to 0 ms or negative, but this comes with a degradation in performance as predicting the future is often a difficult task.

In our experiments, we find that predicting one frame ahead does not dramatically degrade the performance. This is possibly because when predicting one frame ahead and using a loss function like Eq.~(\ref{wav+mag}) which requires training through iSTFT, at frame $t$ we essentially use the input signals up to frame $t$ to predict the sub-frame (marked in the top of Fig.~\ref{overlapdual}(b)) at frame $t$ so that a 2 ms algorithmic latency (i.e., the length of the sub-frame) can be achieved\footnote{In other words, our DNN model in this case can fully observe the mixture signals of the sub-frame at frame $t$, and it has the opportunity to correct at frame $t$ the errors made when predicting at frame $t-1$ the (then in the future) sub-frame at frame $t$. This could be the key reason why predicting one frame ahead performs reasonably well in our experiments.}. We find that it is then important to use the rectangular window as the analysis window and that using tapering-shaped windows produces noticeable artifacts near the boundary of each frame. If we use an analysis window which significantly tapers the right end of the input signal, such as the Tukey or sqrtHann window, the DNN model would have difficulty predicting the signals at the right boundary of the sub-frame at frame $t$, because the input information especially near the right boundary would be lost due to the tapering.

Performance degradation is however significant when predicting two frames ahead (i.e., 4 ms in the future), and even more so when predicting three, possibly because the model now has to fully predict part of the future signal.
If performance could be improved to the point that three (the value of $\text{oWS}/\text{HS}+1$ in our setup) frames ahead can be accurately predicted, for example via a more powerful DNN architecture, %
the algorithmic latency could be reduced to $-2$ ms.
In this case, the enhancement system could potentially achieve $0$ ms processing latency if the hardware latency of processing each frame can be less than the $2$ ms hop size (which is necessary to maintain real-time processing anyway).

\ZQHL{Note that the typical approach to reduce algorithmic latency is to use smaller window and hop sizes, but this increases the computation due to the increased number of frames, and it cannot deal with unavoidable latency due to, for example, analog-to-digital and digital-to-analog conversions.
Future-frame prediction could deal with both issues by using a larger hop size, predicting future frames, and using a strong DNN.}

In our two-DNN system in Fig.~\ref{system_overview_figure}, only DNN$_2$ can choose to predict future frames and DNN$_1$ always predicts the current frame.

\section{Experimental Setup}\label{experiments}

We evaluate the proposed algorithm on a noisy-reverberant speech enhancement task, using a single microphone or an array of microphones.
This section describes the dataset, benchmark systems, and miscellaneous configurations.

\subsection{Dataset for Noisy-Reverberant Speech Enhancement}

Due to the lack of a widely adopted benchmark for multi-channel noisy-reverberant speech enhancement, we build a custom dataset based on the WSJCAM0 \cite{Robinson1995} and FSD50k \cite{Fonseca2020} corpora.
The clean signals in WSJCAM0 are used as the speech sources.
The corpus contains 7,861, 742, and 1,088 utterances respectively in its training, validation, and test sets.
Using the same split of clean signals as in WSJCAM0, we simulate 39,245 ($\sim$77.7 hours), 2,965 ($\sim$5.6 hours), and 3,260 ($\sim$8.5 hours) noisy-reverberant mixtures as our training, validation, and test sets, respectively.
The noise sources are from the FSD50k dataset, which contains around 50,000 Freesound clips with human-labeled sound events distributed in 200 classes drawn from the AudioSet ontology \cite{F.Gemmeke2017}.
We sample the clips in the development set of FSD50k to simulate the noises for training and validation, and those in the evaluation set to simulate the noises for testing.
Since our task is single-speaker speech enhancement, following \cite{Tzinis2021} we filter out clips containing any sounds produced by humans, based on the provided sound event annotation of each clip.
Such clips have annotations such as \textit{Human\_voice}, \textit{Male\_speech\_and\_man\_speaking}, \textit{Chuckle\_and\_chortle}, \textit{Yell}, etc\footnote{See \url{https://github.com/etzinis/fedenhance/blob/master/fedenhance/dataset \_maker/make\_librifsd50k.py} for the full list.}.
To generate multi-microphone noise signals, for each mixture we randomly sample up to seven noise clips.
\ZQHL{We treat each sampled clip as a point source, and its RIR is simulated by using the image method implemented in the Pyroomacoustics software~\cite{Scheibler2018}.
More specifically, for each mixture, we generate a room with its length drawn from the range $[5, 10]$ m, width from $[5, 10]$ m, and height from $[3, 4]$ m, place a simulated six-microphone uniform circular array with a 20 cm diameter in a position randomly drawn in the room, and place each point source with a direction to the array center randomly sampled from the range $[0, 2\pi]$, with a height randomly drawn from $[0.5, 2.5]$ m, and with the distance between each source and the array center drawn from $[0.75, 2.5]$ m. 
The reverberation time (T60) is drawn from the range $[0.2, 1.0]$ s.
We then convolve each source with its simulated RIR, and summate the convolved signals to create the mixture.}
Following the setup in the FUSS dataset \cite{Wisdom2021}, which is designed for universal sound separation, we consider noise clips as background noises if they are more than 10 seconds long, and as foreground noises otherwise.
Each simulated mixed noise file has one background noise and the rest are foreground noises.
The energy level between the dry background noise and each dry foreground noise is drawn from the range $[-3, 9]$ dB.
Considering that some FSD50k clips contain silence or digital zeros, the energy level is computed by first removing silent segments in each clip, next computing a sample variance from the remaining samples, and then scaling the clips to an energy level based on the sample variances.
After summing up all the spatialized noises, we scale the summated reverberant noise such that the SNR between the target direct-path speech and the summated reverberant noise is equal to a value sampled from the range $[-8, 3]$ dB.
Besides the FSD50k clips, in each mixture we always include a weak, diffuse, stationary air-conditioning noise drawn from the REVERB corpus, where the SNR between the target direct-path speech and the noise is equal to a value sampled from the range $[10, 30]$ dB.

The sampling rate is 16 kHz.

\subsection{Benchmark Systems}\label{benchmarkdescription}

We consider the frame-online Conv-TasNet \cite{Luo2019} as the monaural benchmark system.
Conv-TasNet is an excellent model.
It can achieve enhancement with very low algorithmic latency through its very short window length, using a very small amount of computation.
We considered other monaural time-domain approaches such as \cite{Pandey2020}, which uses window sizes as large as typical STFT window sizes and also leverages overlap-add for signal re-synthesis.
It has the same algorithmic latency as regular STFT-based systems due to the overlap-add.
The proposed dual window size approach can be straightforwardly applied to reduce its latency, 
but the model itself requires drastically more computation than Conv-TasNet, mainly due to its DenseNet modules. %
Another recent study \cite{Nakaoka2021} proposes to use low-overlap window for Wave-U-Net. However, a large window is used and their algorithmic latency is at least 38.4 ms.
We therefore only consider Conv-TasNet as the monaural baseline.
We also considered other frame-online T-F domain models such as the winning solutions in the DNS challenges \cite{K.A.Reddy2020}.
However, they are targeted at teleconferencing scenarios, where a processing latency as large as 40 ms is allowed.
For example, DCCRN \cite{Hu2020} has an algorithmic latency of 62.5 ms and TSCN-PP \cite{Li2021} 20 ms.
In addition, these models share many similarities with our complex T-F domain DNN models and can straightforwardly leverage our proposed techniques to reduce their algorithmic latency.
We therefore do not include them as baselines.

We consider the frame-online multi-channel Conv-TasNet \cite{ZhangJisi2020, Tu2021}, denoted as MC-Conv-TasNet, as the %
main multi-channel baseline.
Compared with monaural Conv-TasNet, MC-Conv-TasNet introduces a spatial encoder in addition to the spectral encoder in the original Conv-TasNet to exploit spatial information.
The spatial embedding produced by the spatial encoder is used as extra features for the network to better mask the spectral embedding.
Following \cite{ZhangJisi2020, Tu2021}, we set the spatial embedding dimension to 60 for two-channel processing and to 360 for six-channel processing.
Note that MC-Conv-TasNet is also the enhancement component within the winning system \cite{Tu2021} of the recent Clarity challenge, a major effort in advancing very low-latency speech enhancement. 

The Conv-TasNet models can be trained by
\begin{align}\label{wav}
\mathcal{L}_{\text{Wav}} = & 
\| \hat{s}_q - s_q \|_1,
\end{align}
where $\hat{s}_q$ denotes the predicted signal, or by Eq.~(\ref{wav+mag}).
The magnitude loss in Eq.~(\ref{wav+mag}) can significantly improve speech intelligibility and quality metrics \cite{Wang2021compensation}.
\ZQHL{We can also train Conv-TasNet with the original SI-SDR loss~\cite{LeRoux2019, Luo2019}.}
Using the notations of Conv-TasNet (in Table I of \cite{Luo2019}), the hyper-parameters are set to $N=512, B=158, S_c=158, H=512, P=3, X=8$, and $R=3$ for the single- and multi-channel Conv-TasNets.
$B$ and $S_c$ here are slightly larger than the default 128 in \cite{Luo2019}, since in multi-channel processing there are additional spatial embeddings concatenated to the 512-dimensional spectral embedding as the input to the TCN module of Conv-TasNet.

\begin{figure}
  \centering  
  \includegraphics[width=8.5cm]{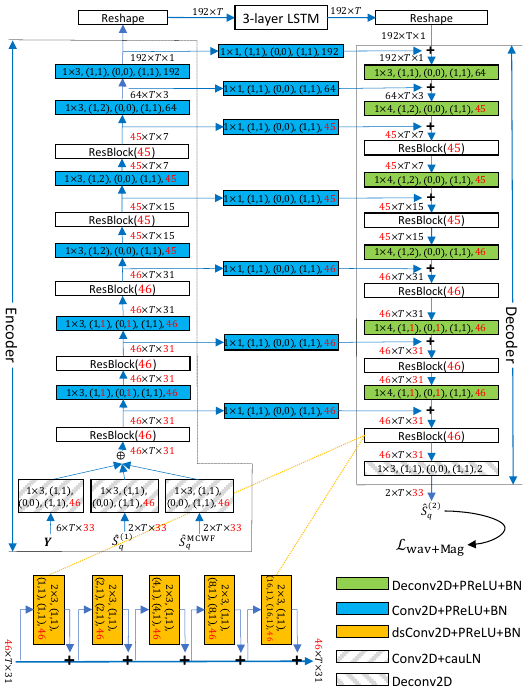} 
  \caption{
  \ZQHL{Network architecture of $\text{DNN}_2$ when using 4 ms iWS without zero padding. We highlight the differences from Fig.~\ref{dnnfigure} in red. Best viewed in color.}
  }
  \label{dnnfigure_4ms_no_padding}\vspace{-0.5cm}
\end{figure}

\ZQHL{
Besides reporting the results of the configuration using 16 ms iWS and 4 ms oWS, we also provide the results of the configurations using 4 ms iWS and 4 ms oWS with or without zero padding.
When using zero padding, we pad each 4 ms windowed signal to 16 ms, use 256-point DFT to extract a 129-dimensional complex spectrum at each frame, and use the same architecture as shown in Fig.~\ref{dnnfigure}
for enhancement.
When not using zero padding, we perform 64-point (i.e., $0.004\times 16,000$) DFT to extract a 33-dimensional complex spectrum at each frame.
Since the input dimension is then lower, we cannot re-use the architecture in Fig.~\ref{dnnfigure} for enhancement.
To deal with this, we design a slightly different architecture shown in Fig.~\ref{dnnfigure_4ms_no_padding}, which uses a similar number of parameters, a similar amount of computation, and the same size of receptive field, compared with Fig.~\ref{dnnfigure}.
The only difference from Fig.~\ref{dnnfigure} is that we use more input and output channels in the convolutional blocks in the first several layers of the encoder and in the last several layers of the decoder, since there are fewer frequencies.
}

\subsection{Miscellaneous Configurations}

The MCWF beamforming filter is updated at each frame.
We pad 
$(\text{iWS}-\text{HS})$ ms of zero samples at the beginning of each mixture.
Without the padding, the algorithmic latency for the starting samples would be higher.

For metric computation, we always use the target direct-path signal as the reference.
It is obtained by setting the T60 parameter to zero when generating RIRs.
Our main evaluation metric is the scale-invariant signal-to-distortion ratio (SI-SDR) \ZQHL{in decibel (dB)} \cite{LeRoux2019}, which measures the quality of time-domain sample-level predictions.
We also report extended short-time objective intelligibility (eSTOI) \cite{H.Taal2011} and perceptual evaluation of speech quality (PESQ) scores.
For PESQ, narrow-band MOS-LQO scores based on the ITU P.862.1 standard \cite{P862.1} are reported using the \textit{python-pesq} toolkit\footnote{\url{https://github.com/ludlows/python-pesq}, v0.0.2}.

For the DNN models, we use the \textit{ptflops} toolkit\footnote{\url{https://github.com/sovrasov/flops-counter.pytorch}} to count the number of floating-point operations (FLOPs) to process a 4-second mixture.
When reporting the FLOPs of an STFT-based system, we summate the DNN FLOPs and the FLOPs of beamforming, STFT, and iSTFT.
Note that two FLOPs is roughly equivalent to one multiply–accumulate operation.

\ZQHL{The number of parameters in each model is reported in millions (M), and the FLOPs in giga-operations (G).}

\begin{table*}[t]
\scriptsize
\centering
\captionsetup{justification=centering}
\sisetup{table-format=2.2,round-mode=places,round-precision=2,table-number-alignment = center,detect-weight=true,detect-inline-weight=math}
\caption{\small{\textsc{\#params (M), FLOPs (G), SI-SDR }(dB), \textsc{PESQ, and eSTOI (\%) Results for Monaural Enhancement.}}}\vspace{-0.1cm}
\label{results1ch}
\setlength{\tabcolsep}{2.5pt}
\begin{tabu}{
c
l
c
c
ccccc
S[table-format=2.1,round-precision=1]
S[table-format=2.2,round-precision=2]
S[table-format=2.1,round-precision=1]
S[table-format=2.1,round-precision=1]
S[table-format=1.2,round-precision=2]!{\enspace}
S[table-format=2.1,round-precision=1]
}
\toprule
& & DNN$_1$ & Window & & & & & Last DNN predicts & {Algorithmic} & & & & & \\
Entry & Systems & Loss & type & \#DFT & {iWS} & {oWS} & {HS} & \#frames ahead & {latency (ms)} & \#params & {FLOPs} & {SI-SDR} & {PESQ} & {eSTOI} \\

\midrule

0 & Unprocessed & {-} & {-} & {-} & {-} & {-} & {-} & {-} & {-} & {-} & {-} & -6.2 & 1.44 & 41.1 \\

\midrule

1a & DNN$_1$ & RI+Mag & Tukey & 256 & 16 & 4 & 2 & 0 & 4 & 2.32 & 27.75668 & 2.75605 & 1.849317 & 66.9980 \\
1b & DNN$_1$ & Wav+Mag & Tukey & 256 & 16 & 4 & 2 & 0 & 4 & 2.32 & 27.75668 & 2.889300 & 1.91038 & 68.377  \\
1c & DNN$_1$ & Wav+Mag & Tukey & 256 & 16 & 4 & 2 & 1 & 2 & 2.32 & 27.75668 & 2.185 & 1.79204 & 64.8963 \\

\midrule

2a & DNN$_1$ & Wav+Mag & Rect & 256 & 16 & 4 & 2 & 0 & 4 & 2.32 & 27.75668 & 2.8341 & 1.8997 & 68.235 \\
2b & DNN$_1$ & Wav+Mag & Rect & 256 & 16 & 4 & 2 & 1 & 2 & 2.32 & 27.75668 & 2.5297 & 1.84575 & 66.637 \\
2c & DNN$_1$ & Wav+Mag & Rect & 256 & 16 & 4 & 2 & 2 & 0 & 2.32 & 27.75668 & -3.6493 & 1.7138 & 62.202 \\
2d & DNN$_1$ & Wav+Mag & Rect & 256 & 16 & 4 & 2 & 3 & -2 & 2.32 & 27.75668 & -5.45738 & 1.626974 & 58.7145 \\

\midrule

3a & Conv-TasNet \cite{Luo2019} & Wav & - & - & 4 & 4 & 2 & 0 & 4 & 6.18 & 29.350856 & 2.2581 & 1.5752 & 61.7139 \\
3b & Conv-TasNet \cite{Luo2019} & Wav+Mag & - & - & 4 & 4 & 2 & 0 & 4 & 6.18 & 29.350856 & 2.1584 & 1.7841 & 65.729 \\
3c & Conv-TasNet \cite{Luo2019} & Wav+Mag & - & - & 4 & 4 & 1 & 0 & 4 & 6.18 & 54.4938 & 2.418 & 1.8326 & 66.68 \\
3d & Conv-TasNet \cite{Luo2019} & Wav+Mag & - & - & 2 & 2 & 1 & 0 & 2 & 6.14 & 52.207439 & 2.181 & 1.7743 & 65.078 \\

\midrule

\ZQHLrow 4a & Conv-TasNet \cite{Luo2019} & SI-SDR & - & - & 4 & 4 & 2 & 0 & 4 & 6.18 & 29.350856 & 2.197 & 1.702 & 61.551 \\
\ZQHLrow 4b & Conv-TasNet \cite{Luo2019} & SI-SDR & - & - & 4 & 4 & 1 & 0 & 4 & 6.18 & 54.4938 & 2.2376 & 1.6678 & 60.848 \\
\ZQHLrow 4c & Conv-TasNet \cite{Luo2019} & SI-SDR & - & - & 2 & 2 & 1 & 0 & 2 & 6.14 & 52.207439 & 2.01380 & 1.6491 & 59.5192 \\

\midrule

\ZQHLrow 5a & DNN$_1$ & Wav+Mag & Rect & 256 & 4 & 4 & 2 & 0 & 4 & 2.32 & 27.728 & 2.59884 & 1.8665 & 67.308 \\
\ZQHLrow 5b & DNN$_1$ & Wav+Mag & Rect & 64 & 4 & 4 & 2 & 0 & 4 & 2.43 & 28.30806 & 2.46150 & 1.849789 & 66.7335 \\

\bottomrule
\end{tabu}\vspace{-0.2cm}
\end{table*}

\begin{table*}[t]
\scriptsize
\centering
\captionsetup{justification=centering}
\sisetup{table-format=2.2,round-mode=places,round-precision=2,table-number-alignment = center,detect-weight=true,detect-inline-weight=math}
\caption{\small{\textsc{SI-SDR} (dB), \textsc{PESQ, and eSTOI (\%) Results Using Various Analysis Windows for Six-Microphone Enhancement.}}}\vspace{-0.1cm}
\label{resultsawin6ch}
\setlength{\tabcolsep}{2.5pt}
\begin{tabular}{
c
l
c
cc
ccc
c
S[table-format=2.1,round-precision=1]
S[table-format=2.2,round-precision=2]
S[table-format=2.1,round-precision=1]
S[table-format=2.1,round-precision=1]
S[table-format=1.2,round-precision=2]!{\enspace}
S[table-format=2.1,round-precision=1]
}
\toprule

      &        & DNN$_1$ & Window &       &       &       &      & Last DNN predicts & {Algorithmic}  & &         &        &         &        \\
Entry & Systems & Loss & type   & \#DFT & {iWS} & {oWS} & {HS} & \#frames ahead   & {latency (ms)} & {\#params} & {FLOPs} & {SI-SDR} & {PESQ} & {eSTOI} \\

\midrule

0 & Unprocessed & - & - & - & - & - & - & - & {-} & {-} & {-} & -6.2 & 1.44 & 41.1 \\

\midrule

1a & DNN$_1$ & RI+Mag & sqrtHann & 256 & 16 & 4 & 2 & 0 & 4 & 2.33 &  28.316568 & 5.8787 & 2.16640 & 75.67 \\
1b & DNN$_1$ & RI+Mag & {AsqrtHann} & 256 & 16 & 4 & 2 & 0 & 4 & 2.33 & 28.316568 & 5.96521 & 2.13058 & 75.9366 \\
1c & DNN$_1$ & RI+Mag & Rect & 256 & 16 & 4 & 2 & 0 & 4 & 2.33 & 28.316568 & 5.991293 & 2.131858 & 75.148 \\
1d & DNN$_1$ & RI+Mag & Tukey & 256 & 16 & 4 & 2 & 0 & 4 & 2.33 & 28.316568 & 6.01940 & 2.16264 & 76.0204 \\

\midrule

2a & DNN$_1$ & Wav+Mag & sqrtHann & 256 & 16 & 4 & 2 & 0 & 4 & 2.33 & 28.316568 & 6.40326 & 2.252724 & 77.2638 \\
2b & DNN$_1$ & Wav+Mag & {AsqrtHann} & 256 & 16 & 4 & 2 & 0 & 4 & 2.33 & 28.316568 & 6.33866 & 2.25233 & 77.1877 \\
2c & DNN$_1$ & Wav+Mag & Rect & 256 & 16 & 4 & 2 & 0 & 4 & 2.33 & 28.316568 & 6.19791 & 2.2300008 & 76.7019 \\
2d & DNN$_1$ & Wav+Mag & Tukey & 256 & 16 & 4 & 2 & 0 & 4 & 2.33 & 28.316568 & 6.376042 & 2.26325 & 77.25547 \\
\bottomrule
\end{tabular}\vspace{-0.4cm}
\end{table*}

\section{Evaluation Results}\label{results}

Tables~\ref{results1ch}, \ref{results6ch} and \ref{results2ch} report the results of one-, six- and two-microphone enhancement, respectively.
In each table, we provide the algorithmic latency of each model, along with the number of model parameters and FLOPs.
When comparing the results, we always 
take into account the algorithmic latency, the amount of computation, \ZQHL{and the model size}.

\subsection{Comparison of Loss Functions}\label{compareloss}

We observe that training through the proposed overlap-add procedure using the Wav+Mag loss function in Eq.~(\ref{wav+mag}) leads to clear improvement over using the RI+Mag loss in Eq.~(\ref{ri+mag}), which 
does not train through the signal re-synthesis procedure.
This can be observed from entries 1a vs.\ 1b in Table~\ref{results1ch}, 1a-1d vs.\ 2a-2d in Table~\ref{resultsawin6ch}, and 1a vs.\ 1b in Table~\ref{results6ch} and \ref{results2ch}.
Using the Wav+Mag loss in Eq.~(\ref{wav+mag}) rather than the Wav loss in Eq.~(\ref{wav}) dramatically improves Conv-TasNet's scores on PESQ and STOI (see 3a vs.\ 3b in Table~\ref{results1ch}, and 5a vs.\ 5b in \ref{results6ch} and \ref{results2ch}).
This aligns with our findings in \cite{Wang2021compensation}, which only deals with offline enhancement.

\begin{table*}[t]
\scriptsize
\centering
\captionsetup{justification=centering}
\sisetup{table-format=2.2,round-mode=places,round-precision=2,table-number-alignment = center,detect-weight=true,detect-inline-weight=math}
\caption{\small{\textsc{\#params (M), FLOPs (G), SI-SDR} (dB), \textsc{PESQ, and eSTOI (\%) Results on Six-Microphone Enhancement.}}}\vspace{-0.1cm}
\label{results6ch}
\setlength{\tabcolsep}{2.5pt}
\begin{tabu}{
c
l
c
c
cc
ccc
c
S[table-format=2.1,round-precision=1]
S[table-format=2.2,round-precision=2]
S[table-format=2.1,round-precision=1]
S[table-format=2.1,round-precision=1]
S[table-format=1.2,round-precision=2]!{\enspace}
S[table-format=2.1,round-precision=1]
}
\toprule

      &        & DNN$_1$ & DNN$_2$ & Window &       &       &       &      & Last DNN predicts & {Algorithmic}  & &         &        &         &        \\
Entry & Systems & Loss   & Loss    & type   & \#DFT & {iWS} & {oWS} & {HS} & \#frames ahead   & {latency (ms)} & \#params & {FLOPs} & {SI-SDR} & {PESQ} & {eSTOI} \\

\midrule

0 & Unprocessed & - & - & - & - & - & - & {-} & - & {-} & {-} & {-} & -6.2 & 1.44 & 41.1 \\

\midrule

1a & DNN$_1$ & RI+Mag & - & Tukey & 256 & 16 & 4 & 2 & 0 & 4 & 2.33& 28.316568 & 6.01940 & 2.16264 & 76.0204 \\
1b & DNN$_1$ & Wav+Mag & - & Tukey & 256 & 16 & 4 & 2 & 0 & 4 & 2.33 & 28.316568 & 6.376042 & 2.26325 & 77.25547 \\
1c & DNN$_1$ & Wav+Mag & - & Tukey & 256 & 16 & 4 & 2 & 1 & 2 & 2.33 & 28.316568 & 5.20235 & 2.09659 & 73.98183 \\

\hdashline

2a & DNN$_1$+MCWF+DNN$_2$ & RI+Mag & Wav+Mag & Tukey & 256 & 16 & 4 & 2 & 0 & 4 & 4.67 & 57.1338 & 7.8954 & 2.698490 & 83.9732 \\
2b & DNN$_1$+MCWF+DNN$_2$ & RI+Mag & Wav+Mag & Tukey & 256 & 16 & 4 & 2 & 1 & 2 & 4.67 & 57.1338 & 6.837408 & 2.47482 & 80.7017 \\
\ZQHLrow 2c & DNN$_1$+MCWF & RI+Mag & - & Tukey & 256 & 16 & 4 & 2 & 0 & 4 & 2.33 & 28.557 & 3.371393 & 1.81485 & 62.7034 \\

\midrule

3a & DNN$_1$ & Wav+Mag & - & Rect & 256 & 16 & 4 & 2 & 0 & 4 & 2.33 & 28.316568 & 6.19791 & 2.2300008 & 76.7019 \\
3b & DNN$_1$ & Wav+Mag & - & Rect & 256 & 16 & 4 & 2 & 1 & 2 & 2.33 & 28.316568 & 5.8899 & 2.20116 & 76.1948 \\
3c & DNN$_1$ & Wav+Mag & - & Rect & 256 & 16 & 4 & 2 & 2 & 0 & 2.33 & 28.316568 & -2.0777 & 1.94296 & 69.9531 \\
\hdashline
4a & DNN$_1$+MCWF+DNN$_2$ & RI+Mag & Wav+Mag & Rect & 256 & 16 & 4 & 2 & 0 & 4 & 4.67 & 57.1338 & 7.76829 & 2.68052 & 83.6202 \\
4b & DNN$_1$+MCWF+DNN$_2$ & RI+Mag & Wav+Mag & Rect & 256 & 16 & 4 & 2 & 1 & 2 & 4.67 & 57.1338 & 7.0890829 & 2.50083 & 80.9806 \\
4c & DNN$_1$+MCWF+DNN$_2$ & RI+Mag & Wav+Mag & Rect & 256 & 16 & 4 & 2 & 2 & 0 & 4.67 & 57.1338 & -1.1426105 & 2.2378 & 76.6883 \\
4d & DNN$_1$+MCWF+DNN$_2$ & RI+Mag & Wav+Mag & Rect & 256 & 16 & 4 & 2 & 3 & -2 & 4.67 & 57.1338 & -2.776644 & 2.12138 & 73.4678 \\
\ZQHLrow 4e & DNN$_1$+MCWF & RI+Mag & - & Rect & 256 & 16 & 4 & 2 & 0 & 4 & 2.33 & 28.557 & 2.8685 & 1.751798 & 59.943 \\

\midrule

5a & MC-Conv-TasNet \cite{Tu2021, ZhangJisi2020} & Wav & - & - & - & 4 & 4 & 2 & 0 & 4 & 6.37 & 30.14188 & 5.4543 & 1.95941 & 73.238 \\
5b & MC-Conv-TasNet \cite{Tu2021, ZhangJisi2020} & Wav+Mag & - & - & - & 4 & 4 & 2 & 0 & 4 & 6.37 & 30.14188 & 5.166 & 2.2353 & 76.352 \\
5c & MC-Conv-TasNet \cite{Tu2021, ZhangJisi2020} & Wav+Mag & - & - & - & 4 & 4 & 1 & 0 & 4 & 6.37 & 56.07585 & 5.6833 & 2.32869 & 77.895 \\
5d & MC-Conv-TasNet \cite{Tu2021, ZhangJisi2020} & Wav+Mag & - & - & - & 2 & 2 & 1 & 0 & 2 & 6.27 & 53.23467 & 5.6747 & 2.29389 & 77.265 \\

\midrule

\ZQHLrow 6a & MC-Conv-TasNet \cite{Tu2021, ZhangJisi2020} & SI-SDR & - & - & - & 4 & 4 & 2 & 0 & 4 & 6.37 & 30.14188 & 5.496117 & 2.06129 & 72.90753 \\
\ZQHLrow 6b & MC-Conv-TasNet \cite{Tu2021, ZhangJisi2020} & SI-SDR & - & - & - & 4 & 4 & 1 & 0 & 4 & 6.37 & 56.07585 & 6.0305 & 2.13153 & 74.542 \\
\ZQHLrow 6c & MC-Conv-TasNet \cite{Tu2021, ZhangJisi2020} & SI-SDR & - & - & - & 2 & 2 & 1 & 0 & 2 & 6.27 & 53.23467 & 6.177458 & 2.126679 & 74.35870 \\

\midrule

\ZQHLrow 7a & DNN$_1$ & Wav+Mag & - & Rect & 256 & 4 & 4 & 2 & 0 & 4 & 2.33 & 28.207 & 6.02228 & 2.19740 & 76.091 \\
\ZQHLrow 7b & DNN$_1$+MCWF+DNN$_2$ & RI+Mag & Wav+Mag & Rect & 256 & 4 & 4 & 2 & 0 & 4 & 4.67 & 57.0201 & 7.308123 & 2.577617 & 82.254962 \\
\ZQHLrow 7c & DNN$_1$+MCWF & RI+Mag & - & Rect & 256 & 4 & 4 & 2 & 0 & 4 & 2.33 & 28.4401 & 2.07075 & 1.7705 & 59.1379 \\
\hdashline
\ZQHLrow 8a & DNN$_1$ & Wav+Mag & - & Rect & 64 & 4 & 4 & 2 & 0 & 4 & 2.43 & 28.487 & 6.196301 & 2.1925 & 75.912794 \\
\ZQHLrow 8b & DNN$_1$+MCWF+DNN$_2$ & RI+Mag & Wav+Mag & Rect & 64 & 4 & 4 & 2 & 0 & 4 & 4.87 & 57.1668 & 7.4331 & 2.58563 & 82.1746 \\
\ZQHLrow 8c & DNN$_1$+MCWF & RI+Mag & - & Rect & 64 & 4 & 4 & 2 & 0 & 4 & 2.43 & 28.5468 & 1.9054 & 1.71016 & 56.82972 \\

\bottomrule
\end{tabu}\vspace{-0.4cm}
\end{table*}

\begin{table*}[t]
\scriptsize
\centering
\captionsetup{justification=centering}
\sisetup{table-format=2.2,round-mode=places,round-precision=2,table-number-alignment = center,detect-weight=true,detect-inline-weight=math}
\caption{\small{\textsc{\#params (M), FLOPs (G), SI-SDR} (dB), \textsc{PESQ, and eSTOI (\%) Results on Two-Microphone Enhancement.}}}\vspace{-0.1cm}
\label{results2ch}
\setlength{\tabcolsep}{2.5pt}
\begin{tabu}{
c
l
c
c
cc
ccc
c
S[table-format=2.1,round-precision=1]
S[table-format=2.2,round-precision=2]
S[table-format=2.1,round-precision=1]
S[table-format=2.1,round-precision=1]
S[table-format=1.2,round-precision=2]!{\enspace}
S[table-format=2.1,round-precision=1]
}
\toprule

      &        & DNN$_1$ & DNN$_2$ & Window &       &       &       &      & Last DNN predicts & {Algorithmic}  &         &          &        & \\
Entry & Systems & Loss    & Loss & type   & \#DFT & {iWS} & {oWS} & {HS} & \#frames ahead   & {latency (ms)} & \#params & {FLOPs} & {SI-SDR} & {PESQ} & {eSTOI} \\

\midrule

0 & Unprocessed & - & - & - & - & - & - & - & - & {-} & {-} & {-} & -6.2 & 1.44 & 41.1 \\

\midrule

1a & DNN$_1$ & RI+Mag & - & Tukey & 256 & 16 & 4 & 2 & 0 & 4 & 2.32 & 27.8626 & 4.02344 & 1.97191 & 70.768 \\
1b & DNN$_1$ & Wav+Mag & - & Tukey & 256 & 16 & 4 & 2 & 0 & 4 & 2.32 & 27.8626 & 4.153133 & 2.04713 & 72.1385 \\
1c & DNN$_1$ & Wav+Mag & - & Tukey & 256 & 16 & 4 & 2 & 1 & 2 & 2.32 & 27.8626 & 3.2806 & 1.92112 & 68.678 \\
\hdashline
2a & DNN$_1$+MCWF+DNN$_2$ & RI+Mag & Wav+Mag & Tukey & 256 & 16 & 4 & 2 & 0 & 4 & 4.66 & 56.1238 & 4.750017 & 2.17424 & 74.46770 \\
2b & DNN$_1$+MCWF+DNN$_2$ & RI+Mag & Wav+Mag & Tukey & 256 & 16 & 4 & 2 & 1 & 2 & 4.66 & 56.1238 & 4.0779 & 2.06351 & 72.1747 \\
\ZQHLrow 2c & DNN$_1$+MCWF & RI+Mag & - & Tukey & 256 & 16 & 4 & 2 & 0 & 4 & 2.33 & 27.899 & -1.07165 & 1.57963 & 48.445 \\

\midrule

3a & DNN$_1$ & Wav+Mag & - & Rect & 256 & 16 & 4 & 2 & 0 & 4 & 2.33 & 27.8626 & 4.25184 & 2.0581 & 72.587 \\
3b & DNN$_1$ & Wav+Mag & - & Rect & 256 & 16 & 4 & 2 & 1 & 2 & 2.33 & 27.8626 & 3.6793 & 1.9704 & 70.1381 \\
3c & DNN$_1$ & Wav+Mag & - & Rect & 256 & 16 & 4 & 2 & 2 & 0 & 2.33 & 27.8626 & -3.18776 & 1.80565 & 65.501 \\
\hdashline
4a & DNN$_1$+MCWF+DNN$_2$ & RI+Mag & Wav+Mag & Rect & 256 & 16 & 4 & 2 & 0 & 4 & 4.66 & 56.1238 & 4.794650 & 2.181987 & 74.8409 \\
4b & DNN$_1$+MCWF+DNN$_2$ & RI+Mag & Wav+Mag & Rect & 256 & 16 & 4 & 2 & 1 & 2 & 4.66 & 56.1238 & 4.2887 & 2.090557 & 72.5485 \\
4c & DNN$_1$+MCWF+DNN$_2$ & RI+Mag & Wav+Mag & Rect & 256 & 16 & 4 & 2 & 2 & 0 & 4.66 & 56.1238 & -2.4606050 & 1.89324 & 68.2596 \\
4d & DNN$_1$+MCWF+DNN$_2$ & RI+Mag & Wav+Mag & Rect & 256 & 16 & 4 & 2 & 3 & -2 & 4.66 & 56.1238 & -4.081644 & 1.811331 & 65.7793 \\
\ZQHLrow 4e & DNN$_1$+MCWF & RI+Mag & - & Rect & 256 & 16 & 4 & 2 & 0 & 4 & 2.33 & 27.899 & -1.1952 & 1.5543 & 47.3021 \\

\midrule

5a & MC-Conv-TasNet \cite{Tu2021, ZhangJisi2020} & Wav & - & - & - & 4 & 4 & 2 & 0 & 4 & 6.19 & 29.42116 & 3.6275 & 1.729968 & 66.9558 \\
5b & MC-Conv-TasNet \cite{Tu2021, ZhangJisi2020} & Wav+Mag & - & - & - & 4 & 4 & 2 & 0 & 4 & 6.19 & 29.42116 & 3.56102 & 1.9950 & 71.1390 \\
5c & MC-Conv-TasNet \cite{Tu2021, ZhangJisi2020} & Wav+Mag & - & - & - & 4 & 4 & 1 & 0 & 4 & 6.19 & 54.63441 & 3.82133 & 2.04237 & 71.968 \\
5d & MC-Conv-TasNet \cite{Tu2021, ZhangJisi2020} & Wav+Mag & - & - & - & 2 & 2 & 1 & 0 & 2 & 6.15 & 52.31717 & 3.7618 & 2.01509 & 71.4074 \\

\midrule

\ZQHLrow 6a & MC-Conv-TasNet \cite{Tu2021, ZhangJisi2020} & SI-SDR & - & - & - & 4 & 4 & 2 & 0 & 4 & 6.19 & 29.42116 & 3.5708 & 1.849667 & 66.859 \\
\ZQHLrow 6b & MC-Conv-TasNet \cite{Tu2021, ZhangJisi2020} & SI-SDR & - & - & - & 4 & 4 & 1 & 0 & 4 & 6.19 & 54.63441 & 3.6546 & 1.8144 & 66.4413 \\
\ZQHLrow 6c & MC-Conv-TasNet \cite{Tu2021, ZhangJisi2020} & SI-SDR & - & - & - & 2 & 2 & 1 & 0 & 2 & 6.15 & 52.31717 & 3.57600 & 1.79164 & 65.642 \\

\midrule

\ZQHLrow 7a & DNN$_1$ & Wav+Mag & - & Rect & 256 & 4 & 4 & 2 & 0 & 4 & 2.32 & 27.8119 & 3.8932 & 1.99509 & 71.033 \\
\ZQHLrow 7b & DNN$_1$+MCWF+DNN$_2$ & RI+Mag & Wav+Mag & Rect & 256 & 4 & 4 & 2 & 0 & 4 & 4.66 & 56.068 & 4.32181 & 2.118584 & 73.52839 \\
\ZQHLrow 7c & DNN$_1$+MCWF & RI+Mag & - & Rect & 256 & 4 & 4 & 2 & 0 & 4 & 2.32 & 27.8485 & -1.62329 & 1.563 & 47.240 \\
\hdashline
\ZQHLrow 8a & DNN$_1$ & Wav+Mag & - & Rect & 64 & 4 & 4 & 2 & 0 & 4 & 2.43 & 28.33 & 3.99185 & 2.00612 & 71.45493 \\
\ZQHLrow 8b & DNN$_1$+MCWF+DNN$_2$ & RI+Mag & Wav+Mag & Rect & 64 & 4 & 4 & 2 & 0 & 4 & 4.87 & 56.821 & 4.4603 & 2.0974 & 73.293 \\
\ZQHLrow 8c & DNN$_1$+MCWF & RI+Mag & - & Rect & 64 & 4 & 4 & 2 & 0 & 4 & 2.43 & 28.341 & -1.6108 & 1.5353 & 46.01043 \\
 
\bottomrule
\end{tabu}
\vspace{-0.4cm}
\end{table*}

\subsection{Comparison of Analysis Windows}\label{maskingvsmappingresults}

We first look at the case where, at each frame, the DNNs are trained to predict the current frame.
When using the Wav+Mag loss and training through the re-synthesis procedure, from 2a-2d in Table~\ref{resultsawin6ch} we find that using different window functions including sqrtHann, AsqrtHann, rectangular, and Tukey windows does not produce notable differences in performance.
This is likely because, via the training-through procedure, the DNNs could learn to deal with the slight differences in the synthesis windows.
Among all the considered windows, the Tukey window appears slightly better than the others.
This can also be observed from 1a-1d in Table~\ref{resultsawin6ch}.

We then look at the results when using the DNNs to predict one frame ahead, which can reduce the algorithmic latency from 4 to 2 ms.
We observe that using the Tukey window leads to more degradation (see 1b vs.\ 1c in Tables~\ref{results1ch}, \ref{results6ch} and \ref{results2ch} and check the ``Last DNN predicts \#frames ahead'' column) than the rectangular window (see 2a vs.\ 2b in Table~\ref{results1ch}, 3a vs.\ 3b in \ref{results6ch} and \ref{results2ch}).
In the end, the Tukey window leads to worse performance than the rectangular window (see 1c vs.\ 2b in Table~\ref{results1ch}, 1c vs.\ 3b in \ref{results6ch} and \ref{results2ch}).
Predicting two frames ahead, which can reduce the algorithmic latency to 0 ms, does not work very well (see 2a and 2b vs.\ 2c in Table~\ref{results1ch}, and 3a and 3b vs.\ 3c in \ref{results6ch} and \ref{results2ch}), especially in terms of SI-SDR.

\subsection{Effectiveness of Beamforming}\label{compareBeamforming}

Comparing 1b and 2a, and 3a and 4a of Table~\ref{results6ch}, we can see that with a beamformer and a post-filtering network, DNN$_1$+MCWF+DNN$_2$ leads to clear improvements especially on PESQ and eSTOI over using DNN$_1$.
Similar trend is observed from 1b and 2a, as well as 3a and 4a of Table~\ref{results2ch}.

\subsection{Comparison with Conv-TasNet}\label{compareTasNet}

In Tables~\ref{results1ch}, \ref{results6ch}, and \ref{results2ch}, we provide the results obtained by single- or multi-channel Conv-TasNet \cite{Luo2019, Tu2021}.
We experiment with 4/2, 4/1, and 2/1 ms window/hop sizes.
Their algorithmic latencies are respectively 4, 4, and 2 ms, and the latter two approximately double the amount of computation used by the first one due to their reduced hop size.
When using the Wav+Mag loss, we found in all the tables that using 4/1 ms window/hop sizes yields consistently better enhancement scores than the other two, possibly because of its higher frame overlap.
\ZQHL{By comparing 3a-3d with 4a-4c of Table~\ref{results1ch}, and 5a-5d with 6a-6c of Tables~\ref{results6ch} and \ref{results2ch}, we observe that training Conv-TasNet using the original SI-SDR loss (1) does not always improve the performance over using Wav, (2) does not always surpass the SI-SDR performance of using Wav+Mag, and (3) does not produce better PESQ and eSTOI scores than Wav+Mag. We therefore choose Wav+Mag as the default loss for Conv-TasNet for subsequent experiments.}

Let us first look at Table~\ref{results1ch}, the monaural results.
Our models use fewer parameters than Conv-TasNet (i.e., 2.32 vs.\ 6.18 M).
When the algorithmic latency is restrained to 4 ms, our system 1b (and 2a) produces better enhancement performance not only than 3b, using a similar number of FLOPs, but also than the best Conv-TasNet model 3c, using around half of the FLOPs.
When the algorithmic latency is limited to 2 ms, the proposed 2b, which predicts one frame ahead, shows better scores than 3d, using around half of the FLOPs.

We now look at Table~\ref{results6ch}.
At 4 ms algorithmic latency, 1b and 3a show slightly better (or comparable in some metrics) results than 5b, using a similar number of FLOPs.
The DNN$_1$+MCWF+DNN$_2$ model contains two DNNs and hence at least doubles the amount of computation of DNN$_1$.
At 4 ms algorithmic latency, our systems in 2a and 4a show clearly better enhancement scores than 5c, using a comparable number of FLOPs.
In 4b, DNN$_2$ predicts one frame ahead and reduces the algorithmic latency to 2 ms. The enhancement scores are clearly better than those in 5d, which also have an algorithmic latency of 2 ms, again using a similar number of FLOPs. Indeed, 4b uses two DNNs, each operating at a hop size of 2 ms, and 5d only uses one DNN but the DNN operates at a hop size of 1 ms.
The comparison between 4b and 5d suggests a new and promising way of achieving speech enhancement with a very low algorithmic latency.
Earlier studies like time-domain methods \cite{Luo2019, Luo2020} tend to use very small window and hop sizes to reduce the algorithmic latency and improve the performance, but this significantly increases the amount of computation due to an increased number of frames.
Differently, we could use larger window and hop sizes (and hence fewer frames) together with more powerful DNN models (such as the proposed two-DNN system with a beamformer in between, which uses more computation at each frame), and at the same time use the proposed future-frame prediction technique to reduce the algorithmic latency.

Similar trends as in Table~\ref{results6ch} can be observed in \ref{results2ch} for two-microphone enhancement.

These comparisons suggest that we can achieve reasonably good enhancement with an algorithmic latency as low as 2 ms in the STFT domain, and that operating in the STFT domain 
may rival or even outperform processing in the time domain for speech enhancement with very low algorithmic latency.

\subsection{Towards Zero Processing Latency}\label{compareTasNet}

In systems 2c and 2d of Table~\ref{results1ch} and 4c and 4d of Tables~\ref{results6ch} and \ref{results2ch}, we train our DNNs to predict two or three frames ahead.
This further reduces the algorithmic latency at the cost of a degradation in performance, compared with the case when we predict one frame ahead.
The degradation is particularly large for SI-SDR, likely because predicting the phase of future frames is difficult.
PESQ and eSTOI, which are less influenced by phase, maintain a decent level of performance, even rivaling at 0 ms algorithmic latency in the six-microphone case with a single-DNN system or an MC-Conv-TasNet system with 4 ms algorithmic latency (see 4c vs.\ 1b and 5c in Table~\ref{results6ch}).
One notable advantage of predicting three frames ahead is that the enhancement system could potentially have a zero processing latency, if the hardware is powerful enough such that the hardware latency can be less than the 2 ms hop size.

\ZQHL{
\subsection{Comparison with Using Equal iWS and oWS}\label{compare_with_regular}

In Tables~\ref{results1ch}, \ref{results6ch}, and \ref{results2ch}, we provide the results of the configuration using 4 ms iWS and 4 ms oWS with and without zero padding (see the last paragraph of Section~\ref{benchmarkdescription} for a description of the setup of this comparison).
When using the DNN$_1$ approach, by comparing 2a, 5a, and 5b in Table~\ref{results1ch}, and by comparing 3a, 7a, and 8a in Tables~\ref{results6ch} and \ref{results2ch}, we observe that using 16 ms iWS produces slight but consistent improvements.
When using the DNN$_1$+MCWF+DNN$_2$ approach, 4a also shows slightly but consistently better performance than 7b and 8b in Tables~\ref{results6ch} and \ref{results2ch}.
}

\section{Conclusion}\label{conclusion}

We have adapted a dual window size approach for deep learning based speech enhancement with very low algorithmic latency in the STFT domain.
Our approach can easily integrate complex T-F domain DNNs with frequency-domain beamforming to achieve better enhancement, without introducing additional algorithmic latency.
A future-frame prediction technique is proposed to further reduce the algorithmic latency.
Evaluation results on a simulated speech enhancement task in noisy-reverberant conditions demonstrate the effectiveness of our algorithms, and show that our STFT-based system can work well at an algorithmic latency as low as 2 ms.
The proposed algorithms can be straightforwardly utilized by, or modified for, many T-F domain or time-domain 
speech separation systems to reduce their algorithmic latency.

The major limitation of our current study comes from the assumption that each frame can be processed within the hop time by hardware in an online streaming setup.
This assumption may not be realistic for edge devices, such as standalone hearing aids with limited computing capability, or even for modern GPUs, unless there is careful design
that can enable the system to achieve frame-by-frame online processing, especially for heavy-duty DNN models.
An ideal speech enhancement system would have a small number of trainable parameters and require a small amount of run-time memory and computation, at the same time achieving high enhancement performance with very low processing latency.
A practical system will likely have to strike a trade-off among these goals, and requires good engineering skills.
Our current study focuses on improving enhancement performance and achieving very low algorithmic latency.
Moving forward, we will consider (a) reducing the DNN complexity and using lightweight DNN blocks \cite{Luo2021GC3}; (b) pruning DNN connections \cite{Tan2021} or quantizing DNN weights \cite{Kim2018}; (c) reducing frequency resolution by using a shorter analysis window; and (d) performing less frequent updates of the beamforming filters.

\bibliographystyle{IEEEtran}
\bibliography{references.bib}

\begin{IEEEbiography}[{\includegraphics[width=1in,height=1.25in,clip,keepaspectratio]{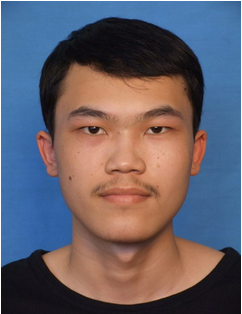}}]{Zhong-Qiu Wang} is a Postdoctoral Research Associate at Carnegie Mellon University, PA, USA. He received the B.E.\ degree in computer science and technology from Harbin Institute of Technology, Harbin, China, in 2013, and the M.S.\ and Ph.D.\ degrees in computer science and engineering from The Ohio State University, Columbus, OH, USA, in 2017 and 2020, respectively. He was a visiting research scientist at MERL, where he worked on deep learning based source separation. His research interests include computer audition, machine hearing, microphone array processing, speech separation, robust automatic speech recognition, machine learning, and deep learning.
\end{IEEEbiography}

\begin{IEEEbiography}
[{\includegraphics[width=1in,height=1.25in,clip,keepaspectratio]{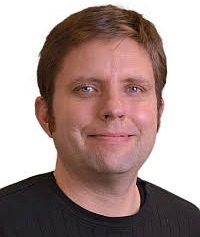}}]{Gordon Wichern}
is a Principal Research Scientist at Mitsubishi Electric Research Laboratories (MERL) in Cambridge, Massachusetts. He received his B.Sc.\ and M.Sc.\ degrees from Colorado State University in electrical engineering and his Ph.D.\ from Arizona State University in electrical engineering with a concentration in arts, media and engineering, where he was supported by a National Science Foundation (NSF) Integrative Graduate Education and Research Traineeship (IGERT) for his work on environmental sound recognition.  He was previously a member of the research team at iZotope, inc.\ where he focused on applying novel signal processing and machine learning techniques to music and post production software, and a member of the Technical Staff at MIT Lincoln Laboratory where he worked on radar signal processing.  His research interests include audio, music, and speech signal processing, machine learning, and psychoacoustics.
\end{IEEEbiography}

\begin{IEEEbiography}
[{\includegraphics[width=1in,height=1.25in,clip,keepaspectratio]{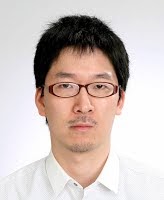}}]{Shinji Watanabe} (Senior Member, IEEE) received the B.S., M.S., and Ph.D. (Dr. Eng.) degrees from Waseda University, Tokyo, Japan, in 1999, 2001, and 2006, respectively. He was the Research Scientist with NTT Communication Science Laboratories, Kyoto, Japan, from 2001 to 2011, Visiting Scholar with Georgia institute of technology, Atlanta, GA, USA, in 2009, and the Senior Principal Research Scientist with Mitsubishi Electric Research Laboratories, Cambridge, MA, USA, from 2012 to 2017. He has authored or authored more than 300 papers in peer-reviewed journals and conferences. His research interests include automatic speech recognition, speech enhancement, spoken language understanding, and machine learning for speech and language processing. He was the recipient of the several awards including the Best Paper Award from the IEEE ASRU in 2019. He was the Associate Editor for the IEEE TRANSACTIONS ON AUDIO SPEECH AND LANGUAGE PROCESSING, and was/has been the Member of several technical committees including the IEEE Signal Processing Society Speech and Language Technical Committee and Machine Learning for Signal Processing Technical Committee.
\end{IEEEbiography}

\begin{IEEEbiography}[{\includegraphics[width=1in,height=1.25in,clip,keepaspectratio]{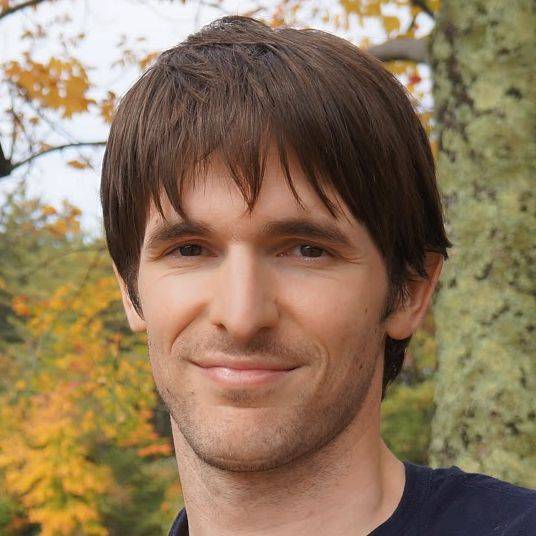}}]{Jonathan Le Roux} is a Senior Principal Research Scientist and the Speech and Audio Senior Team Leader at Mitsubishi Electric Research Laboratories (MERL) in Cambridge, Massachusetts. He completed his B.Sc.\ and M.Sc.\ degrees in Mathematics at the Ecole Normale Sup\'erieure (Paris, France), his Ph.D.\ degree at the University of Tokyo (Japan) and the Universit\'e Pierre et Marie Curie (Paris, France), and worked as a postdoctoral researcher at NTT’s Communication Science Laboratories from 2009 to 2011. His research interests are in signal processing and machine learning applied to speech and audio. He has contributed to more than 130 peer-reviewed papers and 25 granted patents in these fields. He is a founder and chair of the Speech and Audio in the Northeast (SANE) series of workshops, and a Senior Member of the IEEE.
\end{IEEEbiography}

\end{document}